\documentclass[useAMS,usenatbib]{mn2e}

\usepackage{epsfig}
\usepackage{graphics}
\usepackage[usenames]{color}

\usepackage{amssymb}
\usepackage{amsmath}
\usepackage{times}

\newcommand{\be}{\begin{equation}}
\newcommand{\ee}{\end{equation}}
\newcommand{\bdm}{\begin{displaymath}}
\newcommand{\edm}{\end{displaymath}}
\newcommand{\bea}{\begin{eqnarray}}
\newcommand{\eea}{\end{eqnarray}}

\newcommand{\cf}{\textit{cf.}~}
\newcommand{\ie}{\textit{i.e.}~}
\newcommand{\eg}{\textit{e.g.}~}

\title[GR radiation hydrodynamics of accretion flows]
{General relativistic radiation hydrodynamics
of accretion flows. I: Bondi-Hoyle accretion.}

\author[O. Zanotti, C. Roedig, L. Rezzolla, L. Del Zanna]
       {O. Zanotti$^{1}$\thanks{E-mail: zanotti@aei.mpg.de}, C. Roedig$^{1}$,
L. Rezzolla$^{1,2}$ and L. Del Zanna$^{3}$
\\
$^{1}$Max-Planck-Institut f{\"u}r Gravitationsphysik, Albert Einstein Institut, Am M{\"u}hlenberg 1, 14476 Golm, Germany \\
$^{2}$Department of Physics, Louisiana State University,
   Baton Rouge, LA 70803, USA \\
$^{3}$Dipartimento di Fisica e Astronomia,
Universit\`a di Firenze, Largo E. Fermi 2, 50125 Firenze, Italy}
\begin{document}

\date{}

\pagerange{\pageref{firstpage}--\pageref{lastpage}} \pubyear{2009}

\maketitle

\label{firstpage}

\begin{abstract}
We present a new code for performing general-relativistic
radiation-hydrodynamics simulations of accretion flows onto black
holes.  The radiation field is treated in the optically-thick
approximation, with the opacity contributed by Thomson scattering and
thermal bremsstrahlung. Our analysis is concentrated on a detailed
numerical investigation of hot two-dimensional, Bondi-Hoyle accretion
flows with various Mach numbers. We find significant differences with
respect to purely hydrodynamical evolutions. In particular, once the
system relaxes to a radiation-pressure dominated regime, the accretion
rates become about two orders of magnitude smaller than in the purely
hydrodynamical case, remaining however super-Eddington as are the
luminosities. Furthermore, when increasing the Mach number of the
inflowing gas, the accretion rates become smaller because of the
smaller cross section of the black hole, but the luminosities increase
as a result a stronger emission in the shocked regions. Overall, our
approach provides the first self-consistent calculation of the
Bondi-Hoyle luminosity, most of which is emitted within $r\sim 100 M$
from the black hole, with typical values $L/L_{\rm Edd} \simeq 1-7$,
and corresponding energy efficiencies $\eta_{_{\cal BH}}\sim
0.09-0.5$. 
The possibility of computing luminosities self-consistently has
  also allowed us to compare with the bremsstrahlung luminosity often
  used in modelling the electromagnetic counterparts to supermassive
  black-hole binaries, to find that in the optically-thick regime
  these more crude estimates are about $20$ times larger than our
  radiation-hydrodynamics results.
\end{abstract}



\section{INTRODUCTION}
\label{sec:Introduction}

Numerical relativity faces an embarrassing gap between the accuracy
with which it computes the gravitational-wave emission from the
dynamics of compact objects such as black holes and neutron
stars~\citep{McWilliams2010, Duez:2009yz, Rezzolla:2011,
  Sekiguchi2011} and the very rough estimates of the electromagnetic
emission that can be currently computed with state of the art
numerical codes~\citep{Farris08, Palenzuela2009b, Moesta:2009,
  Bode:2009mt, Zanotti2010}. The strongest limitation preventing a
more realistic description of the emitted electromagnetic radiation is
the modelling of radiative transfer in the gas, which is often
neglected in relativistic calculations in view of the large
computational costs involved. This problem is of course common to a
large class of relativistic simulations, but it becomes particularly
apparent in those cases where an accurate computation of the emitted
luminosity is at least as important as providing a faithful
description of the dynamics. Among such cases, accretion onto compact
objects is perhaps the most important one.

Until a few years ago, the time dependent solution of the relativistic
radiation hydrodynamics equations of accretion flows was performed in
one spatial dimension only and, typically, through Lagrangian
finite-difference schemes or through the so called linearized
block-implicit algorithms. Starting from the pioneering works by
\citet{Gilden1980} and \citet{Vitello1984}, relevant results
concerning spherical accretion onto black holes were obtained with a
Lagrangian code by \citet{Zampieri1996}, who were able to solve the
radiation hydrodynamics equations both in the optically thin and in
the optically thick regime, by means of the projected symmetric
trace-free (PSTF) moment formalism introduced by \cite{Thorne1981} and
subsequently reformulated by~\cite{Rezzolla1994} for spherical
flows. Such formalism provides one of the most accurate approximations
to the solution of the radiation transfer equations, and, in analogy
to what is done in fluid dynamics, it allows to define moments of the
radiation field similarly to how density, momentum and pressure of a
medium are defined as moments of the distribution function. As a
result, instead of following rays, the moment equations are solved
directly with an Eulerian or a Lagrangian code.\footnote{In a
  nonrelativistic context, recent interesting developments have been
  reported by \citet{Petkova2009} and \citet{Petkova2010}, who adopted
  the moment formalism within the SPH Gadget code and used a variable
  Eddington tensor as a closure relation, following the Optically Thin
  Variable Eddington Tensor suggestion of~\citet{Gnedin2001}.}

Despite these initial efforts, the time-dependent solution of the
relativistic radiation-hydrodynamics equations in more than one
spatial dimensions remains very challenging. Nowadays, the
multidimensional numerical codes available can be divided in two major
classes, accounting for, separately, the optically thin regime or the
optically thick one. The former class is mainly focused on providing a
realistic modelling of core-collapse supernovae, by employing
Boltzmann neutrino transport, state-of- the-art neutrino interactions,
and general relativity. Relevant achievements have been obtained over
the years by \citet{Mezzacappa01, Bruenn01, Liebendoerfer01,
  Liebendoerfer05a, Messer2008}, who, among other things, showed the
importance of multidimensional simulations to model the shock revival
via neutrinos in a supernova explosion. In a different physical
context, namely that of accretion discs around black holes, but still
in the optically thin regime, \citet{Noble2009} considered an
approximate treatment in which radiation is described through a loss
term in the energy equation. They used fully relativistic ray-tracing
techniques to compute the luminosity received by distant
observers. For a disc with aspect ratio $H /r\simeq 0.1$ accreting
onto a black hole with spin parameter $a=0.9$, they found a
significant dissipation beyond that predicted by the classical model
by~\citet{Novikov:1973}.

The numerical investigation of the optically thick regime, on the
other hand, has received less attention. The seminal work by
\citet{Hsieh1976} already set the basis for the formulation of the
relativistic radiation-hydrodynamics equations in conservation form
and therefore suitable for an Eulerian numerical implementation. Later
on, interesting advances were obtained
by~\citet{Shapiro96},~\citet{Park2006}
and~\citet{Takahashi2007}. Finally, \citet{Farris08} have shown that
for optically-thick gases and gray-body opacities, the general
relativistic radiation-hydrodynamics equations can indeed be written
in conservation form, thus allowing for the use of numerical methods
based on Riemann solvers that have been successfully adopted by many
relativistic hydrodynamics and magnetohydrodynamics codes. Very
recently, and while this paper was being completed,~\cite{Shibata2011}
have presented a modified truncated moment formalism allowing for the
conservative formulation of the relativistic radiation-hydrodynamics
equations both in the optically thin and in the optically thick
limit. This formulation could represent a major step forward with
respect to present leakage schemes accounting for the free streaming
of radiation~\citep{Sekiguchi2010, Sekiguchi2011}.

In this paper, which is the first of a series, we extend our
\texttt{ECHO} code~\citep{DelZanna2007} by following the strategy
suggested by~\citet{Farris08}, and concentrate on one of 
the simplest accretion flows scenarios, namely: Bondi-Hoyle accretion onto a
black hole. This problem has recently been studied
by~\citet{Farris:2009mt} in the context of merging supermassive black
hole binaries in full general relativity, but neglecting
the back-reaction of radiation onto matter.
By assuming that opacity
is made of contributions by Thomson scattering and thermal
bremsstrahlung, we compute here the luminosity emitted in hot
Bondi-Hoyle accretion onto a black hole. As the flow relaxes
to a radiation-pressure dominated regime, we find significant
differences with respect to purely hydrodynamical evolutions. In
particular, the accretion rates drop of about two orders of magnitude
when compared to the purely hydrodynamical case, remaining however
super-Eddington. Furthermore, we find that larger inflow velocities
lead to smaller accretion rates (because of the smaller cross section
of the black hole) but to larger luminosities (because of the stronger
emission in the shocked regions).

The plan of the paper is the following: We first describe the
numerical methods in
Section~\ref{sec:Relativistic_radiation_hydrodynamics} while the
validation of the code is presented in
Section~\ref{sec:Validation_of_the_code}.
Section~\ref{sec:Bondi-Hoyle_accretion_flows}, on the other hand, is
devoted to radiative Bondi-Hoyle accretion flows and contains the main
results of our work. The conclusions are presented in
Sec.~\ref{sec:conclusions}. We assume a signature $(-,+,+,+)$ for the
spacetime metric and we will use Greek letters (running from $0$ to
$3$) for four-dimensional spacetime tensor components, while Latin
letters (running from $1$ to $3$) will be employed for
three-dimensional spatial tensor components. Moreover, we set $c=1$,
$G=10^{-10}$ and extend the geometric units by setting $m_p/k_B=1$,
where $m_p$ is the mass of the proton, while $k_B$ is the Boltzmann
constant. We have maintained $c$, $G$, and $k_B$ in a explicit form in
those expressions of particular physical
interest. Appendix~\ref{appendixA} describes the extended geometrized
system of units adopted in the code.

\section{Relativistic radiation hydrodynamics}
\label{sec:Relativistic_radiation_hydrodynamics}

\subsection{Covariant formulation}
\label{subsec:covariant_formulation}

The total momentum-energy tensor $T^{\alpha\beta}$ of a fluid
immersed in a radiation field comprises two terms:
$T^{\alpha\beta}=T^{\alpha\beta}_{{\rm m}}+T_{\rm r}^{\alpha\beta}$.
The first one is the ordinary one describing the energy and momentum
of the matter 
\be
T^{\alpha\beta}_{{\rm m}}=\rho h\,u^{\,\alpha}u^{\beta}+pg^{\,\alpha\beta}\,,
\label{eq:T_matter}
\ee
where $g^{\alpha\beta}$ is the spacetime metric tensor, $u^\alpha$ is
the four-velocity of the fluid, while $\rho$, $h=1+\epsilon + p/\rho$,
$\epsilon$ and $p$ are the rest-mass density, the specific enthalpy,
the specific internal energy, and the thermal pressure,
respectively. All of these quantities are measured in the comoving
frame of the fluid. The thermal pressure is related to $\rho$ and
$\epsilon$ through an equation of state (EOS), and we will here
consider an ideal-gas, for which the EOS is expressed as
\be
p=\rho\epsilon(\gamma-1) \,,
\ee
where $\gamma$ is the (constant) adiabatic index of the gas. The
second term describes instead the radiation field and is given
by~\citep{Mihalas84,Shapiro96}
\be
\label{eq:T_radiation}
T_{\rm r}^{\alpha\beta}=\frac{1}{c}\int I_\nu N^\alpha N^\beta d\nu d\Omega \,,
\ee
where $I_\nu=I_\nu(x^\alpha,N^i,\nu)$ is the specific
intensity\footnote{We note that $I_\nu$ is an energy flux per unit
  time, frequency and solid angle, so that in cgs units it has
  dimensions of ${\rm erg \ cm^{-2}\ s^{-1}\ Hz^{-1}\ sr^{-1}}$.}  of
the radiation, $N^\alpha$ is the four-vector defining the photon
propagation direction, $d\nu$ is the infinitesimal frequency and
$d\Omega$ is the infinitesimal solid angle around the direction of
propagation. We recall that the direction of propagation of the photon
is defined as \hbox{$N^\alpha\equiv p^\alpha/h_{\rm Pl}\nu$}, where
$p^\alpha$ is the photon four-momentum, while $h_{\rm Pl}$ and $\nu$
are, respectively, the Planck constant and the photon frequency as
measured in the comoving frame of the fluid. Since the
two terms $\nu d\nu d\Omega$ and $I_\nu/\nu^3$ are relativistic
invariants~\citep{Rybicki_Lightman1986}, their product with the tensor
$p^\alpha p^\beta$ is still a tensor, and indeed it provides the
integrand of Eq.~(\ref{eq:T_radiation}).

In the frame comoving with the fluid, the moments of the radiation
field are the energy density, the radiation flux and the radiation
stress tensor, which are respectively given by
\bea
&&E_{\rm r}=\frac{1}{c}\int I_\nu d\nu d\Omega \,, \\
&&F_{\rm r}^\alpha=h^\alpha_{\ \beta}\int I_\nu d\nu d\Omega N^\beta
\,, \\
&&P^{\alpha\beta}_{\rm r}=\frac{1}{c}\int I_\nu d\nu d\Omega N^\alpha N^\beta
\,,
\eea
where the tensor $h^{\alpha\beta}=g^{\alpha\beta}+u^\alpha
u^\beta$ projects any other tensor into the space orthogonal to
$u^\alpha$, namely $h^{\alpha\beta}u_\alpha=0$.
In terms of such moments the radiation energy-momentum
tensor $T_{\rm r}^{\alpha\beta}$ can be rewritten as \citep{Hsieh1976}
\be
T_{\rm r}^{\alpha\beta}=(E_{\rm r}+{\cal P}_{\rm r}) u^\alpha u^\beta +
F_{\rm r}^\alpha u^\beta+ u^\alpha F_{\rm r}^\beta + {\cal P}_{\rm r}g^{\alpha\beta} \,,
\ee
where $E_{\rm r}$ and ${\cal P}_{\rm r}$ are the radiation energy
density and pressure, respectively. As in~\citet{Farris08}, we make
the additional and strong physical assumption that the radiation is
very close to being isotropic in the comoving frame of the fluid, thus
mimicking the conditions of the optically thick regime. However, while
the radiation pressure is actually set to be ${\cal P}_{\rm r}=E_{\rm
  r}/3$, as the isotropic assumption implies, the radiation flux is
allowed to assume nonvanishing values, although with the constraint
that $F_{\rm r}^i/E_{\rm r}\ll 1$. Hence, the radiation field is only
approximately isotropic.

The full set of equations describing the dynamics of the system is
\bea
\label{0eq:mass}
&&\nabla_{\alpha} (\rho u^{\,\alpha})=0, \\
\label{0eq:momentum}
&&\nabla_{\alpha}T^{\alpha\beta}=0, \\
\label{0eq:rad}
&&\nabla_\alpha T_{\rm r}^{\alpha\beta}=-G^\beta_{\rm r} \,.
\eea

While Eqs.~(\ref{0eq:mass}), and~(\ref{0eq:momentum}) represent the
well known continuity equation and the energy momentum equation,
Eq.~(\ref{0eq:rad}) expresses the evolution of the radiation field,
where $G^\alpha_{\rm r}$ is the radiation four-force density. The
latter depends on the physical interaction between matter and
radiation and is therefore specific to the problem considered. In full
generality this tensor is given by \citep{Mihalas84,Shapiro96}
\be
G^\alpha_{\rm r}=\frac{1}{c}\int(\chi_\nu I_\nu-\eta_\nu)N^\alpha
d\nu d\Omega \,,
\ee
where $\chi_\nu \equiv \chi_\nu^t+\chi_\nu^s$ and $\eta_\nu\equiv
\eta_\nu^t+\eta_\nu^s$ are the total opacity and emissivity
coefficients\footnote{Note that although both are referred to as
  ``coefficients'', $\chi_\nu$ and $\eta_\nu$ have different
  units. The dimensions of $\chi_\nu$ are ${\rm cm^{-1}}$, while those
  of $\eta_\nu$ are ${\rm erg \ cm^{-3} s^{-1} Hz^{-1} sr^{-1}}$.},
each containing a thermal contribution, indicated with the superscript
``t'', and a scattering one, indicated with a superscript ``s''. In
addition, we assume that: \textit{(i)} the scattering is isotropic and
coherent; \textit{(ii)} the thermal emissivity and the thermal opacity
coefficients are related to the Planck function $\tilde{B}_\nu$
through Kirchhoff's law $\eta^t_\nu=\tilde{B}_\nu\chi_\nu^t$;
\textit{(iii)} that electrons and ions are maintained at the same
temperature; \textit{(iv)} the opacity coefficients are independent of
frequency, $\chi_\nu=\kappa_{g}\rho$, where $\kappa_g$ is the
gray-body opacity. The last assumption, in particular, prevents us
from taking into account photoionization effects, which are therefore
not considered in our analysis.

Under these conditions, which are indeed the same considered by
\cite{Farris08}, the radiation four-force can be written in covariant
form as
\be
\label{four_force}
G^\alpha_{\rm r}=\chi^t(E_{\rm r}-4\pi \tilde{B})u^\alpha+
(\chi^t+\chi^s)F_{\rm r}^\alpha\,,
\ee
where $4\pi \tilde{B}=a_{\rm rad}T^4$ is the equilibrium black-body
intensity, with $T$ the temperature of the fluid and $a_{\rm rad}$ is
the radiation constant. The temperature is estimated from the
ideal-gas EOS via the expression
\be
\label{t_estimate}
T=\frac{m_p}{k_B}\frac{p}{\rho}\,,
\ee
where, we recall, $k_B$ is the Boltzmann constant and $m_p$ the
rest-mass of the proton. In this paper we consider the case of
bremsstrahlung opacity~\citep{Rybicki_Lightman1986}
\bea
\label{bremsstrahlung_opacity}
\chi^t_{br}&=&1.7\times 10^{-25}\,T_K^{-7/2}\,Z^2\,n_e\,
n_i  \ \ {\rm cm^{-1}}  \nonumber \\
&=&1.7\times 10^{-25}\,T_K^{-7/2}\, \frac{\rho^2_{{\rm cgs}}}{m_p^2}
 \ \ {\rm cm^{-1}} \,,
\eea
where $n_e$ and $n_i\simeq n_e$ are respectively the number densities
of electrons and ions (protons) expressed in cgs units, while $T_K$ is
the equilibrium temperature of both electrons and protons expressed in
Kelvin. For the scattering opacity we consider the Thomson
scattering opacity and we recall that the Thomson cross sections of
electrons and protons are: $\sigma_{T,e}=6.6524586\times 10^{-25} {\rm
  cm}^2$ and $\sigma_{T,p}=(m_e/m_p)^2\sigma_{T,e}$,
respectively. Hence, the Thomson scattering opacities of electrons and
of protons are given by
\bea
\label{scattering_thompson_e} 
&&\hskip -0.5cm
\chi^s_e=\sigma_{T,e}\, n_e =
\sigma_{T,e}\,\left(\frac{\rho}{m_p}\right) = 0.397726
\,\rho_{\rm{cgs}} \, {\rm   cm^{-1}} \,, \\
\label{scattering_thompson_p}
&&\hskip -0.5cm
\chi^s_p=\sigma_{T,p}\, n_p =
\sigma_{T,p}\,\left(\frac{\rho}{m_p}\right) = 1.17968\times 10^{-7}\,
\rho_{\rm{cgs}} \,{\rm cm^{-1}} \,.\nonumber \\
\eea
We recall that the electron-scattering opacity dominates over
free-free opacity at low densities and high
temperatures~\citep{Harwit1998}, where the interaction between
electrons and ions is weak. 
It is worth stressing that,
because of the assumptions made, an incoherent process such as Compton
scattering, with a cross section that is frequency dependent, cannot
be consistently taken into account and it is therefore neglected.
Finally, as customary, the optical thickness is defined as the line
integral of the opacities between two points in the fluid 
\be
\label{eq:tau}
\tau=\int_0^L (\chi^t + \chi^s) ds \,.
\ee
In practice, we approximate expression~(\ref{eq:tau}) as $\tau\simeq
(\chi^t + \chi^s) L$, with $L$ being a typical length scale of the
problem.

We refer to
Appendix~\ref{appendixA} for a summary about the conversion between
$\rm {cgs}$ and geometrized units.

\subsection{Numerical methods}
\label{subsec:numerical_method}

We solve the equations of general relativistic non-dissipative
radiation hydrodynamics \eqref{0eq:mass}-\eqref{0eq:rad} through a
modified version of the \texttt{ECHO} code~\citep{DelZanna2007}, which
adopts a $3+1$ split of spacetime in which the spacetime is foliated
into non-intersecting space-like hyper-surfaces $\Sigma_t$, defined as
isosurfaces of a scalar time function $t$. Within this approach, the
metric is decomposed according to~\citep{Arnowitt62}
\be
\mathrm{d}s^2 = \! -\alpha^2\mathrm{d}t^2+\gamma_{ij}\,
(\mathrm{d}x^i\!+\beta^i\mathrm{d}t)(\mathrm{d}x^j\!+\beta^j\mathrm{d}t),
\label{eq:adm}
\ee
where $\alpha$ is the lapse function, $\beta^i$ is the shift vector,
$\gamma_{ij}$ is the spatial metric tensor, and
\be
n_{\mu}=-\alpha\nabla_{\mu} t=(-\alpha,0_i),~~~~  (n_{\,\mu} n^{\,\mu}=-1)\,,
\label{eq:n}
\ee
is the future-pointing time-like unit vector normal to the slices
$\Sigma_t$. The observer moving with four-velocity
$n^{\,\mu}=\{1/\alpha,-\,\beta^i/\alpha\}$ is called \emph{Eulerian}
\citep{Smarr78b}. Any vector $V^{\,\mu}$ (or similarly a tensor) may
be projected in its temporal component $V^{\hat{n}}=-n_{\mu}V^{\,\mu}$
and spatial component $_{\perp}V^{\,\mu} = (g^{\,\mu}_{\ \nu} +
n^{\,\mu}n_{\nu}) V^{\nu}$. As a result, any spatial vector $V^\mu$
(or tensor) must necessarily have a vanishing contravariant temporal
component $V^t=0$, whereas its covariant temporal component is
$V_t=g_{\mu t}V^\mu=\beta_i V^i$, in general different from zero.
The $3+1$ splitting procedure just described can be applied to the
vectors and tensor introduced so far to yield
\bea
   u^{\,\alpha} & = & \Gamma\, n^{\,\alpha} + \Gamma\, v^{\,\alpha}\,,
\label{eq:u} \\
T_{\rm m}^{\alpha\beta}& = & W^{\alpha\beta} + S^{\alpha}n^{\beta}+ n^{\,\alpha}S^{\beta} + Un^{\,\alpha}n^{\beta}\,,
\label{eq:T} \\
F_{\rm r}^\alpha & =&\alpha F_{\rm r}^t n^\alpha + f_{\rm r}^\alpha \,,
\label{eq:F_perp} \\
T_{\rm r}^{\alpha\beta} & =& R^{\alpha\beta}_{\rm r} + S_{\rm r}^{\alpha}n^{\beta}+
n^{\,\alpha}S_{\rm r}^{\beta} + U_{\rm r} n^{\,\alpha}n^{\beta}\,,
\label{eq:R}
\eea
where all the tensors $v^\mu$, $W^{\mu\nu}$, $S^\mu$, $f_{\rm r}^\mu$,
$R^{\mu\nu}_{\rm r}$, $S_{\rm r}^\mu$ correspond to the familiar
three-dimensional quantities as measured by the Eulerian observers,
are purely spatial, and have indices that are raised and lowered by
the spatial metric $\gamma_{ij}$. In particular, the newly introduced
quantities are related to the corresponding quantities in the comoving
frame by
\bea
D & \equiv & \rho \Gamma\,,\\
W^{ij} & \equiv & \rho h \Gamma^2v^i\,v^j +p\,\gamma^{ij}\,,
\label{eq:W} \\
S^i & \equiv & \rho h \Gamma^2v^i ,
\label{eq:S} \\
U & \equiv & \rho h \Gamma^2 - p\,,
\label{eq:U} \\
R^{ij}_{\rm r} & \equiv &\frac{4}{3}E_{\rm r} \Gamma^2 v^iv^j
+\Gamma(f_{\rm r}^iv^j+f_{\rm r}^jv^i)+{\cal P}_{\rm r}\gamma^{ij} \,,
\label{eq:Rperp} \\
S_{\rm r}^i & \equiv & \frac{4}{3}E_{\rm r} \Gamma^2v^i + \Gamma(\alpha
F_{\rm r}^t v^i + f_{\rm r}^i) \,,
\label{eq:S_r} \\
U_{\rm r}& \equiv &\frac{4}{3}E_{\rm r} \Gamma^2 + 2\alpha
\Gamma F_{\rm r}^t -\frac{E_{\rm r}}{3} \,.
\label{eq:U_r}
\eea
A few comments about the quantities in the equations above can be
useful. The vectors $v^i$ and $f_{\rm r}^i$ are the velocity and the
radiation flux, respectively, as measured by the Eulerian observers,
while $\Gamma=(1-v^2)^{-1/2}=\alpha u^t$ is the Lorentz factor of the
bulk flow. In particular, the radiation flux vector is $f_{\rm
  r}^i=F_{\rm r}^i+\beta^i F_{\rm r}^t$ where $F_{\rm r}^t$ is
computed from the orthogonality condition $F_{\rm r}^\alpha
u_\alpha=0$ and is given by
\be
F_{\rm r}^t=\frac{v_i F_{\rm r}^i}{\alpha-\beta_i v^i} =
\frac{v_i f_{\rm r}^i}{\alpha}\,.
\ee
It is interesting to note that $U_{\rm r}=R^{\alpha\beta}_{\rm r}
n_\alpha n_\beta$ is the radiation energy density as measured by the
Eulerian observers, in analogy with what happens for the conserved
energy density of the fluid $U$ defined by \eqref{eq:U}.

The general-relativistic radiation-hydrodynamics equations are then
written in the following conservative form
\be
\partial_t\vec{\mathcal{U}} + \partial_i\vec{\mathcal{F}}^i=\vec{\mathcal{S}}\,,
\label{eq:UFS}
\ee
which is appropriate for numerical integration via standard
high-resolution shock-capturing (HRSC) methods developed for the Euler
equations. The conservative variables and the corresponding fluxes in
the $i$-direction are respectively given by
\be
\vec{\mathcal{U}}\equiv\sqrt{\gamma}\left[\begin{array}{c}
D \\ \\ S_j \\ \\U \\ \\ (S_{\rm r})_{j} \\ \\ U_{\rm
  r}
\end{array}\right],~~~
\vec{\mathcal{F}}^i\equiv\sqrt{\gamma}\left[\begin{array}{c}
\alpha v^i D-\beta^i D \\\\
\alpha W^i_{\ j}-\beta^i S_j \\\\
\alpha S^i-\beta^i U \\\\
\alpha R^{ij}_{\rm r}-\beta_i (S_{\rm r})_{j} \\\\
\alpha S^i_{\rm r}-\beta^i U_{\rm r} \\\\
\end{array}\right] ,
\label{eq:fluxes}
\ee
whereas the sources, in any stationary background metric, can be
written as 
\be
\vec{\mathcal{S}} \equiv \sqrt{\gamma}\left[\begin{array}{c}
0 \\  \\
\frac{1}{2}\alpha W^{ik}\partial_j\gamma_{ik}+
S_i\partial_j\beta^i-U\partial_j\alpha+\alpha (G_{\rm r})_j \\ \\
\frac{1}{2}W^{ik}\beta^j\partial_j\gamma_{ik}+{W_i}^j\partial_j\beta^i
-S^j\partial_j\alpha +\alpha^2 G^t_{\rm r} \\ \\
\frac{1}{2}\alpha R^{ik}_{\rm r}\partial_j\gamma_{ik}+
(S_{\rm r})_{i}\partial_j\beta^i-U_{\rm r}\partial_j\alpha-\alpha (G_{\rm r})_j \\ \\
\frac{1}{2}R^{ik}_{\rm r}\beta^j\partial_j\gamma_{ik}+(R_{\rm r})_{\ i}^j\partial_j\beta^i
-S_{\rm r}^j\partial_j\alpha-\alpha^2 G^t_{\rm r}
\end{array}\right],
\ee
where only purely spatial quantities are present. We note that
$\sqrt{\gamma}\equiv \sqrt{-g}/\alpha$ is the determinant of the
spatial metric. In our setup for two dimensional simulations
presented in Sec.~\ref{sec:Bondi-Hoyle_accretion_flows} we assume the
metric given by the Kerr solution with the limiting case of
Schwarzschild metric for vanishing black-hole spins.

The radial numerical grid is discretized by choosing $N_r$ points from
$r_\mathrm{min}$ to $r_\mathrm{max}$, non-uniformly distributed
according to the following scheme
\bea
r_i &=& r_\mathrm{min} + a_1 \tan{(a_2 x_i)} \,, \\
x_i &=&
(\tilde{r}_i-r_\mathrm{min})/(r_\mathrm{max}-r_\mathrm{min})
\,,
\eea
where $a_1=(r_\mathrm{max}-r_\mathrm{min})/a_0$, $a_2=\arctan{a_0}$,
while $\tilde{r}_i$ are the coordinate points of the uniform grid from
$r_\mathrm{min}$ to $r_\mathrm{max}$. In practice, the free parameter
$a_0$ controls the extent to which the grid points of the original
uniform grid are concentrated towards $r_\mathrm{min}$, and we have
chosen $a_0$ in the range $[5-10]$ in most of our simulations. The angular
grid is taken to be uniform.

%
\begin{table*}
\begin{center}
\caption{Description of the initial states in the shock-tube tests
  with radiation field. The different columns refer respectively to:
  the test considered, the radiation constant, the adiabatic index
  and the thermal opacity. Also reported are the rest-mass density,
  pressure, velocity and radiation energy density in the ``left''
  ($L$) and ``right'' ($R$) states.}
\label{tab1}
\begin{tabular}{llllllllllll|}
\hline
\hline
Model & $\gamma$ & $a_{\rm rad}$ & $\kappa_g^t$ & $\rho_L$
& $p_L$ &  $u^x_L$ & $E_{{\rm r},L}$ & $\rho_R$ & $p_R$ &
$u^x_R$ & $E_{{\rm r},R}$    \\

\hline

\texttt{1} & $5/3$ &$1.234\times10^{10}$ &$0.4$ &$1.0$ &$3.0\times10^{-5}$ & $0.015$ & $1.0\times10^{-8}$&$2.4$ & $1.61\times 10^{-4}$  &  $6.25\times10^{-3}$ & $2.51\times 10^{-7}$ \\
\texttt{2} & $5/3$ &$7.812\times10^{4}$ &$0.2$ &$1.0$ &$4.0\times10^{-3}$ & $0.25$ & $2.0\times10^{-5}$&$3.11$ &
$0.04512$  &  $0.0804$ & $3.46\times 10^{-3}$ \\
\texttt{3} & $2$   &$1.543\times10^{-7}$ &$0.3$ &$1.0$ &$60.0$ & $10.0$ & $2.0$&$8.0$ &
$2.34\times 10^3$  &  $1.25$ & $1.14\times 10^3$ \\
\texttt{4} & $5/3$ &$1.388\times10^8$ &$0.08$ & $1.0$ &
$6.0\times10^{-3}$ & $0.69$ &  $0.18$ & $3.65$ & $3.59\times 10^{-2}$  &  $0.189$ & $1.3$ \\

\hline
\hline
\end{tabular}
\end{center}
\end{table*}
%

\begin{figure*}
{\includegraphics[angle=0,width=7.5cm,height=7.5cm]{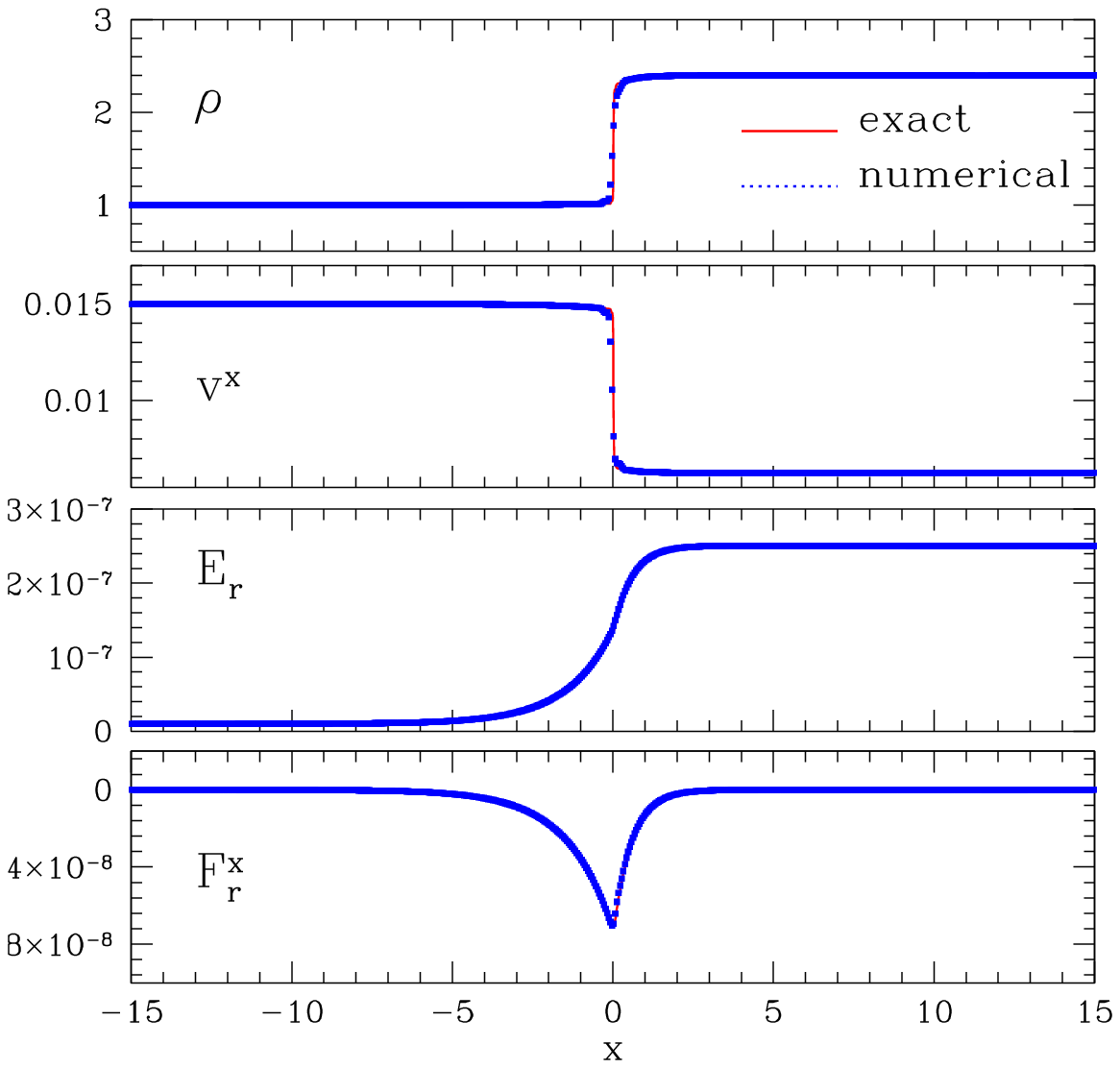}}
\hskip 1.0cm
{\includegraphics[angle=0,width=7.5cm,height=7.5cm]{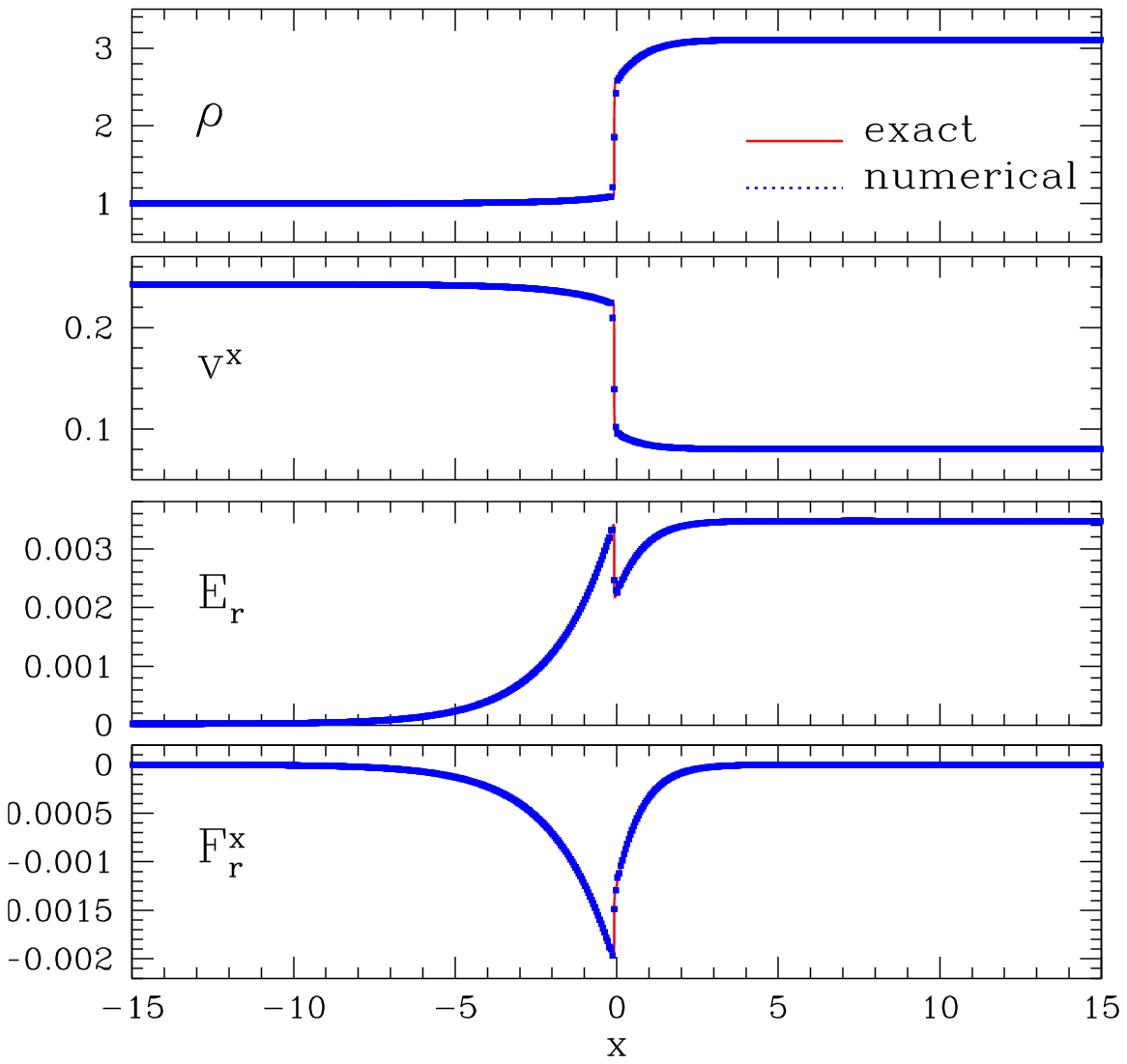}}
\vspace*{-0.0cm}
\caption{Solution of the shock tube test \texttt{1} (left panel) and
  \texttt{2} (right panel) as reported in Table~\ref{tab1}. From top
  to bottom the panels report the rest-mass density, the velocity, the
  radiation energy density and the radiation flux.}
\label{fig:test1-2}
\end{figure*}

The set of hydrodynamics equations is discretized in time with the
method of lines and the evolution is performed with a second-order
modified Euler scheme. A fifth-order finite-difference algorithm based
on an upwind monotonicity-preserving filter is employed for spatial
reconstruction of primitive variables, whereas a two-wave HLL Riemann
solver is used to ensure the shock-capturing properties
(see~\citet{DelZanna2007} for further details). As a final remark we
note that as customary in HRSC methods, we introduce a tenuous and
static ``atmosphere'' in the regions of the fluid where the rest-mass
density falls below a chosen threshold value. When this happens, we
follow the prescription detailed in~\citet{Baiotti04} as far as the
hydrodynamical quantities are concerned, while the primitive variables
of the radiation field are frozen to the values at previous time-step.

\section{Validation of the code}
\label{sec:Validation_of_the_code}
The purely magnetohydrodynamic version of the code has been validated
over the same numerical tests extensively described
in~\citet{DelZanna2007}, obtaining the same convergence properties and
will not be reported here for compactness. For the radiation part of
the code, on the other hand, there are only a few analytic or
semi-analytic tests that can be adopted, as we discuss below.

\subsection{Shock-tube problems}

Considering a flat spacetime, we have followed~\citet{Farris08}, who
proposed and solved four shock-tube tests in which nonlinear
radiation-hydrodynamic waves propagate. The initial states of these
tests are reported in Table~\ref{tab1} and are chosen in such a way
that the discontinuity front at $x=0$ remains stationary, namely it
has zero velocity with respect to the Eulerian observer of the
code. The values of the fluxes, not reported in Table~\ref{tab1}, are
chosen to be two orders of magnitude smaller than the energy density
of the radiation field. In these tests local thermodynamic equilibrium
is assumed at both ends $x=\pm X$, with $X=20$, and this is obtained
by adopting a fictitious value of the radiation constant $a_{\rm
  rad}$, namely $a_{\rm rad}=E_{{\rm r},L}/T_L^4$, which is then used
to compute $E_{{\rm r},R}=a_{\rm rad}T_R^4$ (here the indices $L$ and
$R$ indicate the ``left'' and ``right'' states, respectively). The
scattering opacity $\kappa_g^s$ is set to zero in all of the tests,
while the value of the thermal opacity $\kappa_g^t$ is reported in
Table~\ref{tab1}. 

Each test is evolved in time until stationarity is reached. The
semi-analytic solution that is used for comparison with the numerical
one has been obtained following the strategy by~\citet{Farris08}, and
it implies the solution of the following system of ordinary
differential equations
\[
d_x \mathbf{U}(\mathbf{P}   ) = \mathbf{S}(\mathbf{P}) \,,
\]
where
\[\mathbf{P} = \left(
\begin{array}{c}
\rho\\
P\\
u^x\\
E_{\rm r}\\
F_{\rm r}^x
\end{array}\right)\,,
\qquad
\mathbf{U} = \left(
\begin{array}{c}
\rho u^x\\
T^{0 x} \\
T^{x x}\\
T_{\rm r}^{0 x} \\
T_{\rm r}^{x x}
\end{array}\right)  \,,
\qquad
\mathbf{S} =\left(
\begin{array}{c}
0\\
0\\
0\\
 -G^0_{\rm r}\\
 -G^x_{\rm r}
\end{array}\right)\,.\]

\begin{figure*}
{\includegraphics[angle=0,width=8.0cm,height=7.5cm]{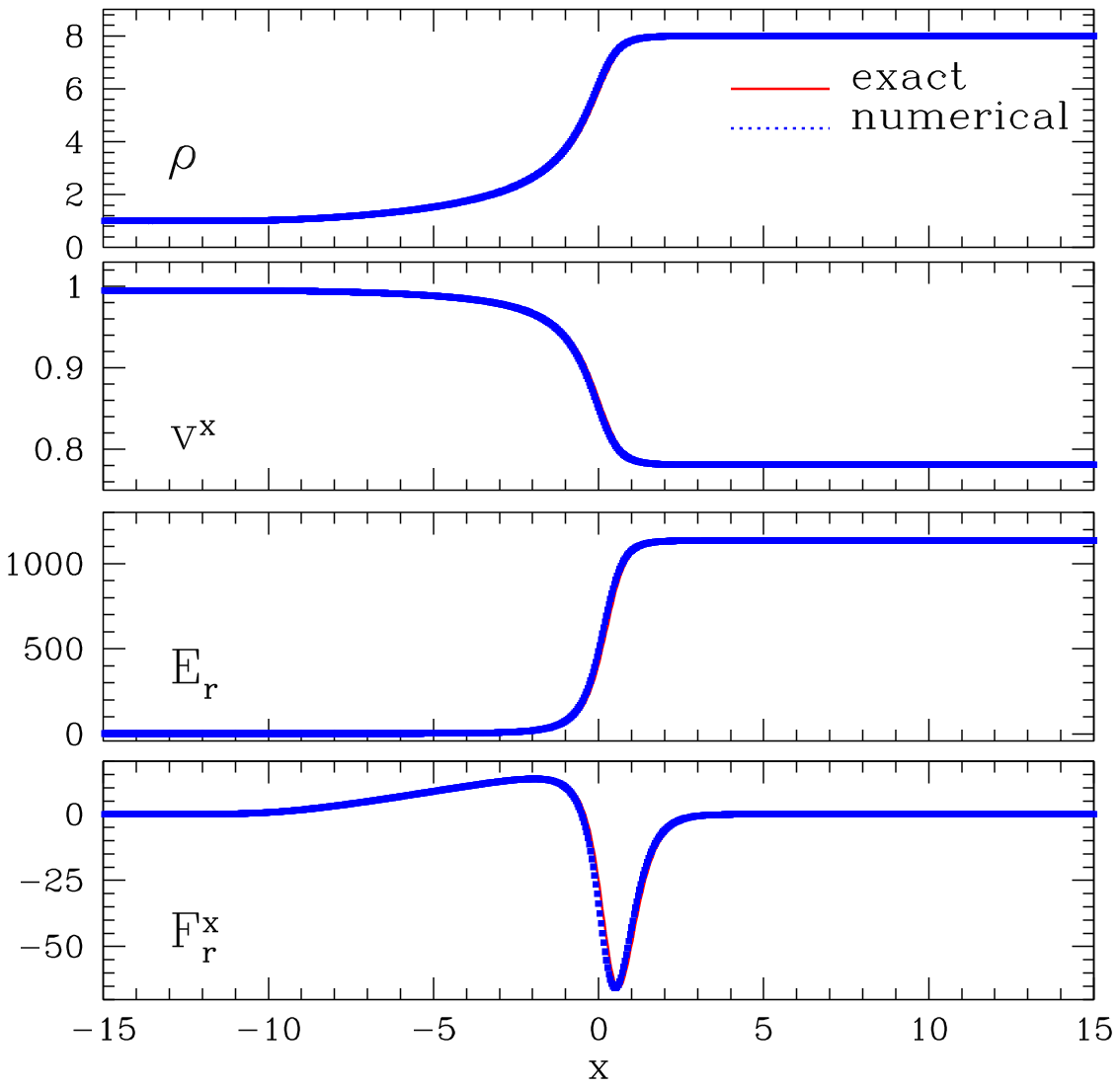}}
\hskip 1.0cm
{\includegraphics[angle=0,width=8.0cm,height=7.5cm]{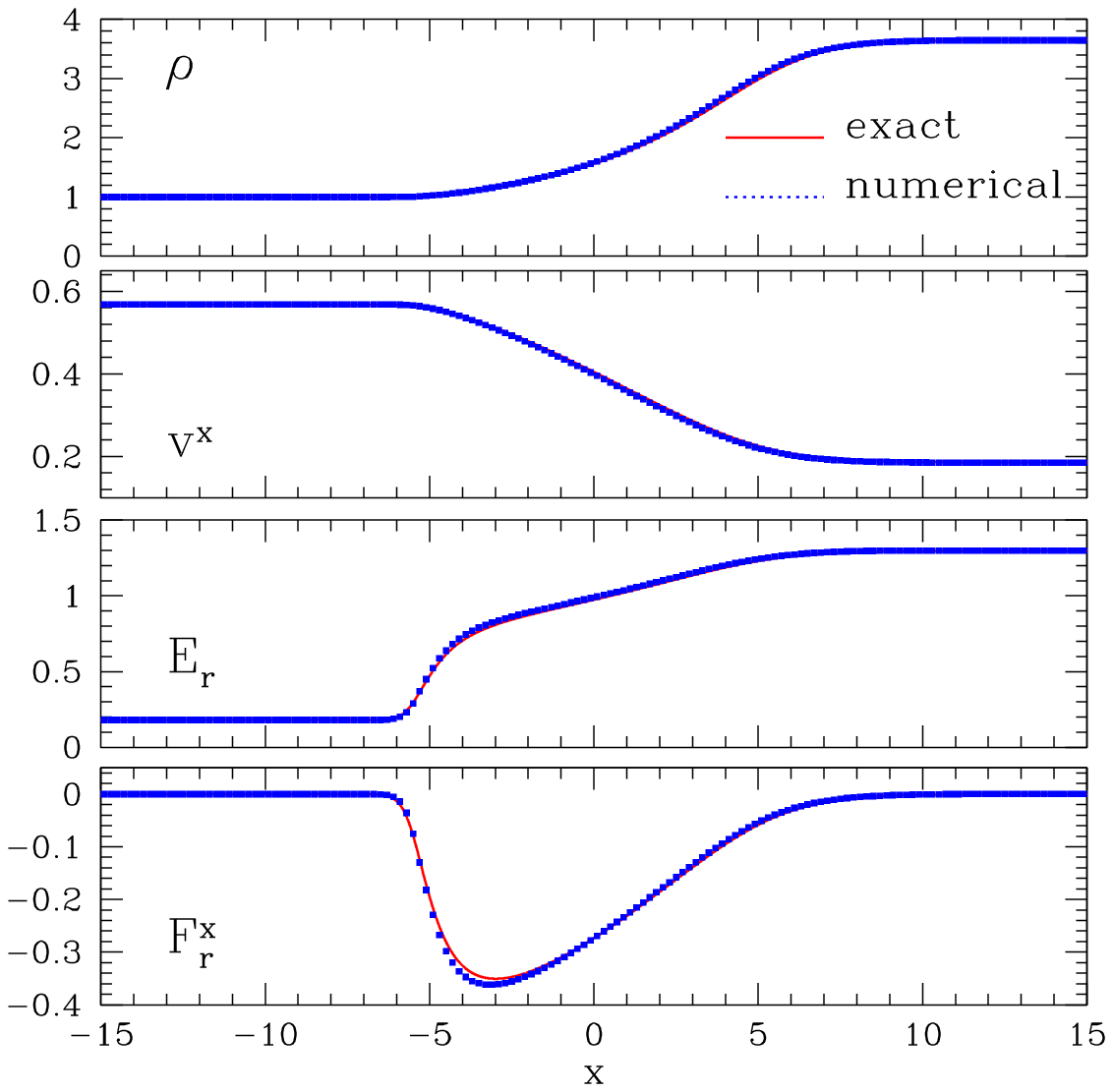}}
\vspace*{-0.0cm}
\caption{Solution of the shock tube test \texttt{3} (left panel) and
  \texttt{4} (right panel) as reported in Table~\ref{tab1}. From top
  to bottom the panels report the rest-mass density, the velocity, the
  radiation energy density and the radiation flux.}
\label{fig:test3-4}
\end{figure*}

Figures~\ref{fig:test1-2} and~\ref{fig:test3-4} show the comparison of
the numerical solution with respect to the semi-analytic one in the
four cases considered, which correspond, respectively, to the
propagation of a nonrelativistic strong shock, of a mildly
relativistic strong shock, of a highly relativistic wave and of a
radiation pressure dominated mildly relativistic wave. In particular,
Fig.~\ref{fig:test1-2} reports the solution for the tests \texttt{1}
and \texttt{2}, which contain a true discontinuity represented by a
shock front, while tests \texttt{3} and \texttt{4} have continuous
configurations and are shown in Fig.~\ref{fig:test3-4}.

The tests have been performed with $N=800$ uniformly spaced grid
points using the MP5 slope limiter described in~\citet{DelZanna2007}
and a HLL Riemann solver. Unlike~\citet{Farris08}, we have not boosted
the solution. This results in a more stringent test for the code to
maintain stationarity and it also explains why the profiles of the vector
quantities, namely the velocity and the radiation flux, do not match
those shown by~\citet{Farris08}. The numerical solution is almost
indistinguishable from the semi-analytic one in all of the profiles
reported in the figures, thus proving the ability of the code in
handling different physical regimes of the radiation field within an
optically-thick approximation.

\section{Bondi-Hoyle accretion flows}
\label{sec:Bondi-Hoyle_accretion_flows}

\subsection{Initial and  boundary conditions}
\label{sec:Initial_conditions}

\begin{table*}
  \caption{Initial models adopted in numerical simulation. From left
    to right the columns report: the name of the model, the asymptotic
    flow velocity $v_{\infty}$, the asymptotic sound speed ${\rm
      c}_{s,\infty}$, the asymptotic Mach number ${\cal M}_{\infty}$,
    the initial temperature, the initial rest-mass density and the
    accretion radius $r_{\rm{a}} \equiv G M/(v_{\infty}^2+{\rm
      c}_{s,\infty}^2)$. Perturbed Bondi-Hoyle solutions are generated
    through injection of low pressure gas as described in the
    text. Each model is evolved until stationarity is reached, and in
    any case up until at least $t=20000\,M$. The adiabatic index was
    set to $\gamma=5/3$ and the black hole spin $a=0$. 
    The mass of the black hole is $M_{BH}=3.6\times
    10^6\, M_{\odot}$ and
    the radial grid
    extends from $r_{\rm {min}}=2.1\,M$ to $r_{\rm{max}}=200\,M$. 
 \label{table:Initial Models}}
\begin{center}
  \begin{tabular}{cccccccc}
    \hline \hline
                Model & $v_{\infty}$ & $c_{s,\infty}$ &
                ${\cal M}_{\infty}$ & $T\ [K]$ & $\rho_{\infty}\ [{\rm cgs}]$& $r_{{\rm a}}\ [M]$ \\
    \hline
$\mathtt{V08.CS07}$   & $0.08$ & $0.07$ & $1.14$ & $3.22\times10^{10}$ & $3.22\times10^{-12}$& $88.5$\\
$\mathtt{V09.CS07}$   & $0.09$ & $0.07$ & $1.28$ & $3.22\times10^{10}$ & $3.22\times10^{-12}$& $76.9$\\
$\mathtt{V10.CS07}$   & $0.10$ & $0.07$ & $1.42$ & $3.22\times10^{10}$ & $3.22\times10^{-12}$& $67.1$\\
$\mathtt{V11.CS07}$   & $0.11$ & $0.07$ & $1.57$ & $3.22\times10^{10}$ & $3.22\times10^{-12}$& $58.8$\\
   \hline
$\mathtt{V07.CS06}$   & $0.07$ & $0.06$ & $1.16$ & $2.36\times10^{10}$ & $4.39\times10^{-12}$& $117.6$\\
$\mathtt{V07.CS07}$   & $0.07$ & $0.07$ & $1.0$ & $3.22\times10^{10}$ & $3.22\times10^{-12}$& $102.0$\\
$\mathtt{V07.CS08}$   & $0.07$ & $0.08$ & $0.77$ & $4.22\times10^{10}$ & $2.45\times10^{-12}$& $88.5$\\
$\mathtt{V07.CS09}$   & $0.07$ & $0.09$ & $0.77$ & $5.35\times10^{10}$ & $1.93\times10^{-12}$& $76.9$\\
    \hline
$\mathtt{p.V09.CS07}$ & $0.09$ & $0.07$ & $1.28$ & $3.22\times10^{9}$  &$3.22\times10^{-12}$& $ 76.9  $\\
$\mathtt{p.V10.CS07}$ & $0.10$ & $0.07$ & $1.42$ & $3.22\times10^{9}$  &$3.22\times10^{-12}$& $67.1$\\
$\mathtt{p.V11.CS07}$ & $0.11$ & $0.07$ & $1.57$ & $3.22\times10^{9}$  &$3.22\times10^{-12}$& $58.8$\\
$\mathtt{p.V18.CS07}$ & $0.18$ & $0.07$ & $2.57$ & $3.22\times10^{9}$  &$3.22\times10^{-12}$& $26.8$\\
   \hline
    \hline
  \end{tabular}
\end{center}
\end{table*}

Our attention is focused on a
Bondi-Hoyle accretion flow onto a black hole of galactic
size with $M_{BH}=3.6\times 10^6\, M_{\odot}$,
that we investigate by
performing numerical simulations on the equatorial plane, \ie $\theta
= \pi/2$. Despite the long history in literature on this type of
accretion (see the review by \citet{Edgar2004}), no stationary
solution for a radiation-hydrodynamics Bondi-Hoyle flow is known, and
which could have been used as suitable initial data. As a result, we let the
code converge to the nearest stationary solution after specifying the
hydrodynamical solution of the Bondi-Hoyle flow, to which we add a
radiation field with uniform and small energy density $E_{\rm r}$.

Most of our discussion hereafter refers to accretion onto
Schwarzschild black holes, although also rotating black holes will be
briefly presented in Sect.~\ref{sec:sbhs}. The code solves therefore
the equations in a general Kerr metric expressed in Boyer-Lindquist
coordinates, so that the initial velocity field, specified in terms of
an asymptotic velocity $v_{\infty}$, is given by~\citep{Font98a}
\begin{eqnarray}
\label{asymtotic_vel_bla}
v^r &=& \sqrt{\gamma^{rr}} v_{\infty} \cos\phi \,,\\
v^{\phi} &=&  -\sqrt{\gamma^{\phi \phi}}  v_{\infty}
\sin\phi \,.
\label{asymtotic_vel_blb}
\end{eqnarray}

These relations guarantee that the velocity of the injected gas at
infinity is parallel to the $x-$direction, while $v^2 \equiv v_iv^i=
v_{\infty}^2$ everywhere in the flow. Other quantities that need to be
set initially are: the asymptotic sound speed $c_{s,\infty}$, and the
asymptotic pressure, from which the asymptotic rest-mass density
$\rho_{\infty}$ follows directly (see values reported in
Table~\ref{table:Initial Models} for all of the models
considered). For all the simulations we will consider a gas of
nonrelativistic electrons and hence with an adiabatic index
$\gamma=5/3$. The velocities used in our models and presented in
Table~\ref{table:Initial Models} are chosen to be sufficiently high so
as to open a shock cone (see details below). Any chosen $v_{\infty}$
implies a restricted range of asymptotic sound speeds, if a reasonable
Mach number should be considered. We remark that our models do not aim
at modelling any specific astrophysical scenario, but rather at
highlighting the role of the back-reaction of the radiation in an
optically thick, relativistic Bondi-Hoyle accretion flow.

Similarly, the radiation field is initialized to a value such that the
radiation temperature $T_{\rm rad}=(E_{\rm r}/a_{\rm
  rad})^{1/4}\approx 1.5\times 10^5 K$. While this may seem an
arbitrary choice, we have verified through a series of numerical
simulations that, on long-term evolutions, the value of the obtained
luminosity is not dependent of this initial choice. The computational
grid consists of $N_r \times N_{\phi}$ numerical cells in the radial
and angular directions, respectively, covering a computational domain
extending from $r_{\rm {min}}=2.1\,M$ to $r_{\rm {max}}=200\,M$ and
from $\phi_{\rm {min}}=0$ to $\phi_{\rm {max}}=2\pi$. For our fiducial
simulation we have chosen $N_r=1536$ and $N_\phi=300$, but have also
verified that the results are not sensitive to the resolution used or
to the location of the outer boundary.

The boundary conditions in the radial direction are such that at the
inner radial grid point we implement inflow boundary conditions by a
simple zeroth-order extrapolation (\ie a direct copy) of all
variables. At the outer radial boundary, on the other hand, we must
distinguish between the upstream region (\ie with
$\pi/2<\phi<3/2\pi$), and the downstream region (\ie with
$-\pi/2<\phi<\pi/2$). In the upstream region we continuously inject
matter with the initial velocity field
of~(\ref{asymtotic_vel_bla})-(\ref{asymtotic_vel_blb}), thus
reproducing a continuous wind at large distances, while in the
downstream region we use outflow boundary conditions. Finally,
symmetric (\ie periodic) boundary conditions are adopted at $\phi=0$.
The simulations are performed with a Courant-Friedrichs-Lewy
coefficient that may vary according to the model and it is typically
in the range $\sim[0.01,0.5]$.

In addition to the ``classical'' Bondi-Hoyle initial data, we will
also consider a set of simulations in which the thermodynamics of the
flow is slightly altered in order to reduce the temperature of the
gas. We denote these models as ``p-models'' in
Table~\ref{table:Initial Models}. In essence, the perturbed
Bondi-Hoyle accretion flows are obtained by injecting gas
 of lower pressure than required by the stationary solution at the upwind boundary,
 with a proportionally reduced radiation energy density $E_{\rm r}$
 (see Sect.~\ref{Perturbed Bondi-Hoyle
  flow} for details).

\subsection{Computation of luminosity}
\label{subsec:luminosity}

The key new quantity that our code allows to compute is the
emitted luminosity. Since the code explicitly calculates the radiation
fluxes $f_{\rm r}^i$ at each time step, we use them to compute the
intrinsic luminosity emitted from the optically thick region as
\be
L=\int_{\Omega} f_{\rm r}^i\, n_i \,d\, S_{opt}\,,
\label{eq:lumexact}
\ee
where $S_{opt}$ is the surface of the volume $\Omega$ enclosing an
optically thick region within the computational domain, while the
scalar product $f_{\rm r}^i\, n_i $ provides the projection of the
local radiation flux onto the normal to the radiating surface.
Because of the nearly isotropic assumption made for the
radiation field and because of
the rough spherical symmetry of the physical system under
consideration, the fluxes in the angular directions are expected to to
be much smaller than the radial ones and to almost cancel. As a
result, and for simplicity, we approximate the scalar product above as
$f_{\rm r}^i\, n_i = f^r_{\rm r}$, thus computing the luminosity as
\be
\label{luminosity}
L= 2\,\sum_{n=1}^{N_\phi} \left[\sqrt{\gamma}\,\left(f^r_{\rm r}\right)_n \,
\Delta \phi_n\right]\vert_{\tau=1}\,,
\ee
where $\Delta \phi_n$ is the angular size of a grid cell and we
perform the surface integral at the radial position of the last
optically-thick surface, \ie where $\tau=1$; the factor $2$ accounts
for both the contributions above and below the equatorial plane. 

The luminosity computed in this way comprises two different
contributions. The first one is an accretion-powered luminosity that
it is directly proportional to the mass-accretion rate $\dot M$
through a relation of the type $L_{\rm acc}=\eta \dot M$, where the
coefficient $\eta$ expresses the efficiency of the conversion of
gravitational binding energy into radiation. The main dissipative
mechanism is provided by compression of the fluid when this has
nonzero thermal conductivity\footnote{We recall that the thermal
  conductivity is related to the opacity and its effects are therefore
  accounted for in our analysis.  For instance, the thermal
  conductivity computed using the ordinary diffusion approximation of
  stellar interiors is given by $\chi_T=(4/3) a_{\rm rad}c
  T^3/\chi^s$~\citep{Schwartz1967}.}. A second contribution to the
total luminosity~\eqref{luminosity} is given by dissipative processes
related to shock heating that, as we will show below, can provide a
considerable contribution to the total emission.

However, since we are dealing with inviscid non-magnetized fluids, the
luminosity~\eqref{luminosity} obviously cannot provide the
contribution coming from dissipative processes driven by viscosity (of
whatever origin), and that can be a significant part of the
accretion-powered luminosity in a realistic accretion scenario. We
recall, for instance, that in the classical Shakura-Sunyaev thin-disc
model the main dissipative mechanism comes from the viscous stress
tensor, directly proportional to the total pressure via the ``alpha''
parameter.  Similarly, in spherical accretion, a realistic viscous
fluid with nonzero bulk viscosity will produce a viscous dissipation
adding to the one coming from the fluid compression. In summary: in
realistic accretion scenarios one should expect that {\em both}
thermal conductivity and viscosity act as transport coefficients of
dissipative processes and lead to contributions to the emitted
luminosity. In our treatment, however, only the effects of the former
one can be accounted for. Hereafter, the luminosities and the
accretion rates will be reported in Eddington units, \ie $L_{\rm
  Edd}=4\pi G M m_p c / \sigma_{T,e} \simeq 1.26\times
10^{38}\,(M/M_\odot)\, {\rm erg \ s^{-1}}$, $\dot M_{\rm Edd}=L_{\rm
  Edd}/c^2\simeq 1.39\times 10^{17}\,(M/M_\odot)\, {\rm g \ s^{-1}}$.
See also~(\ref{Edd_geo}) for the Eddington luminosity in the
geometrized units of the code.

\begin{figure*}
\centering
{\includegraphics[angle=0,width=8.0cm,height=7.3cm]{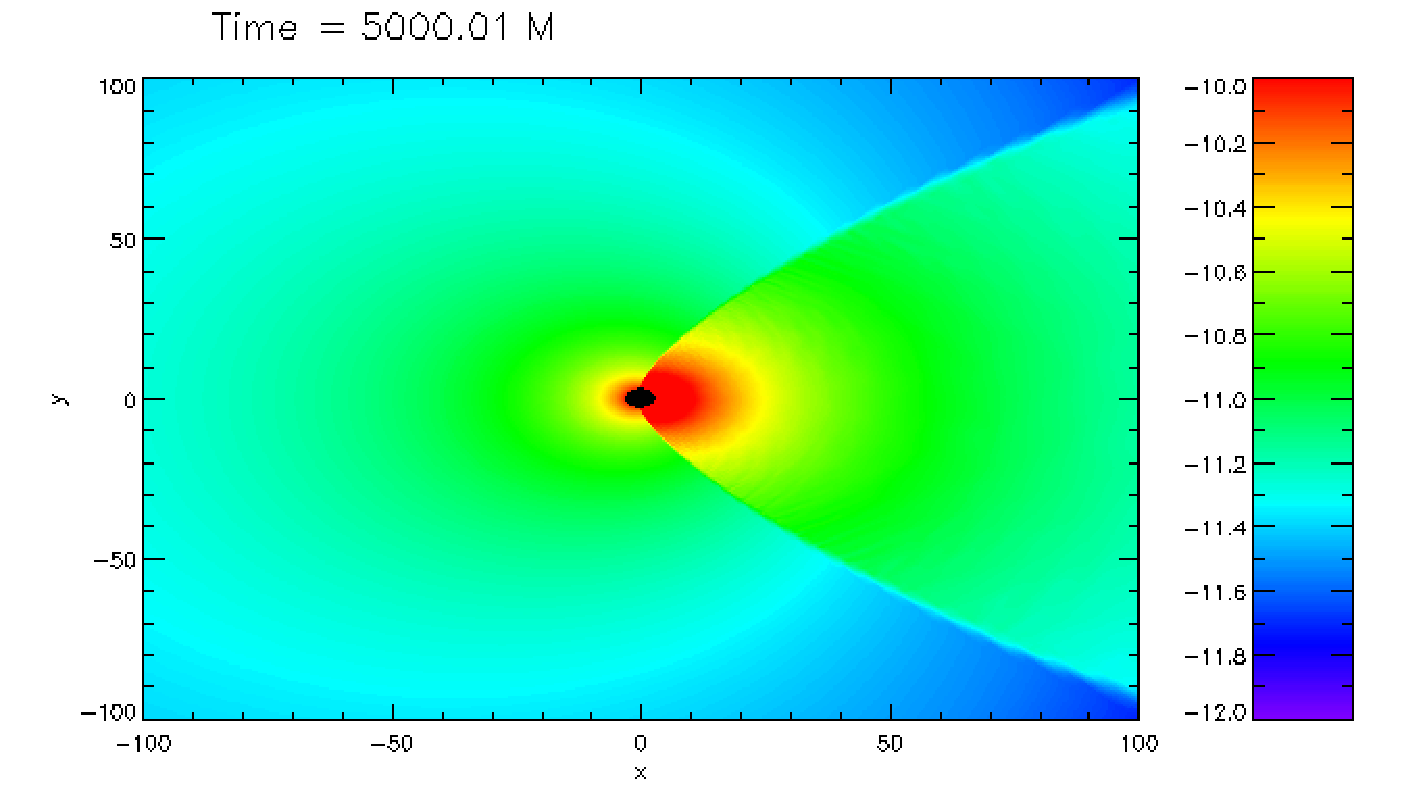}}
{\includegraphics[angle=0,width=8.0cm,height=7.3cm]{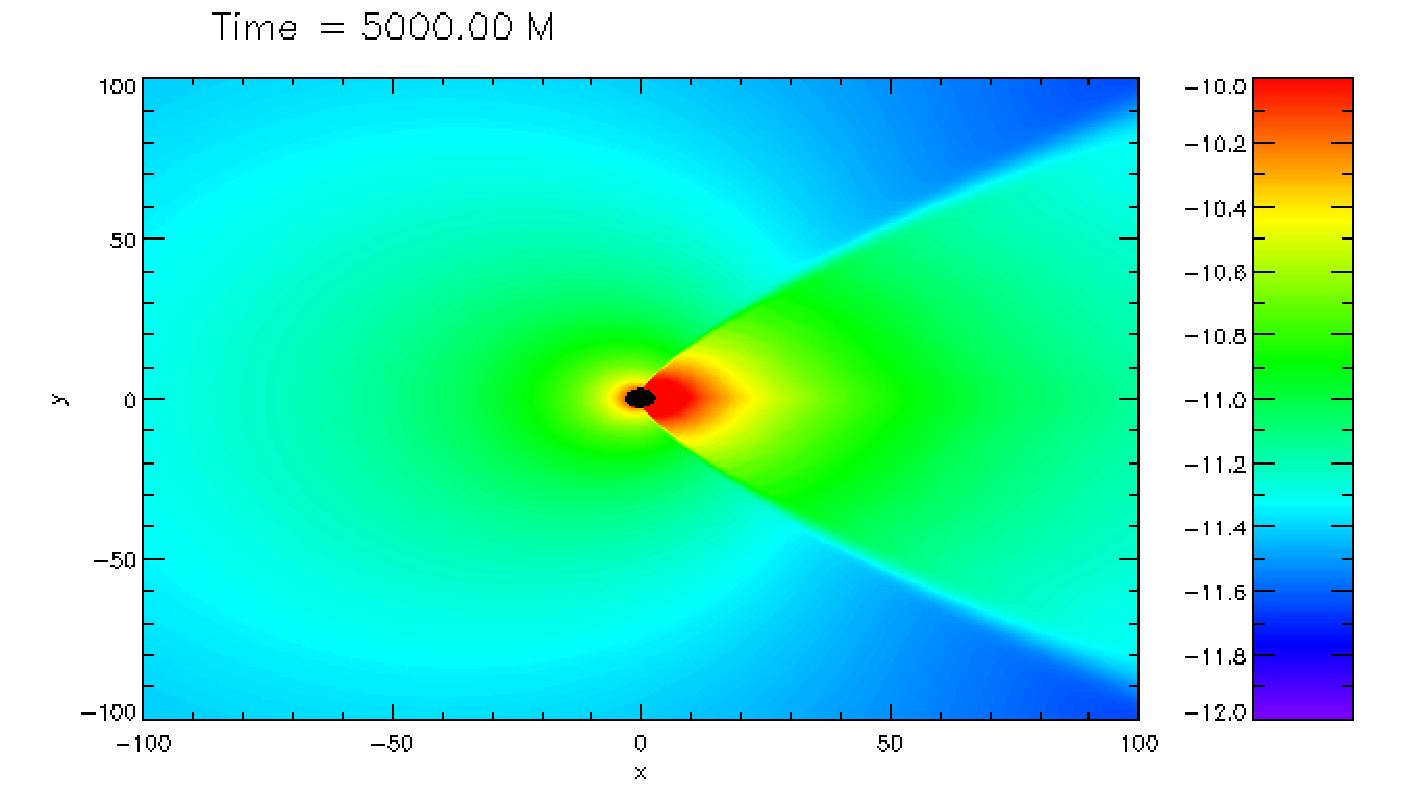}}
{\includegraphics[angle=0,width=8.0cm,height=7.3cm]{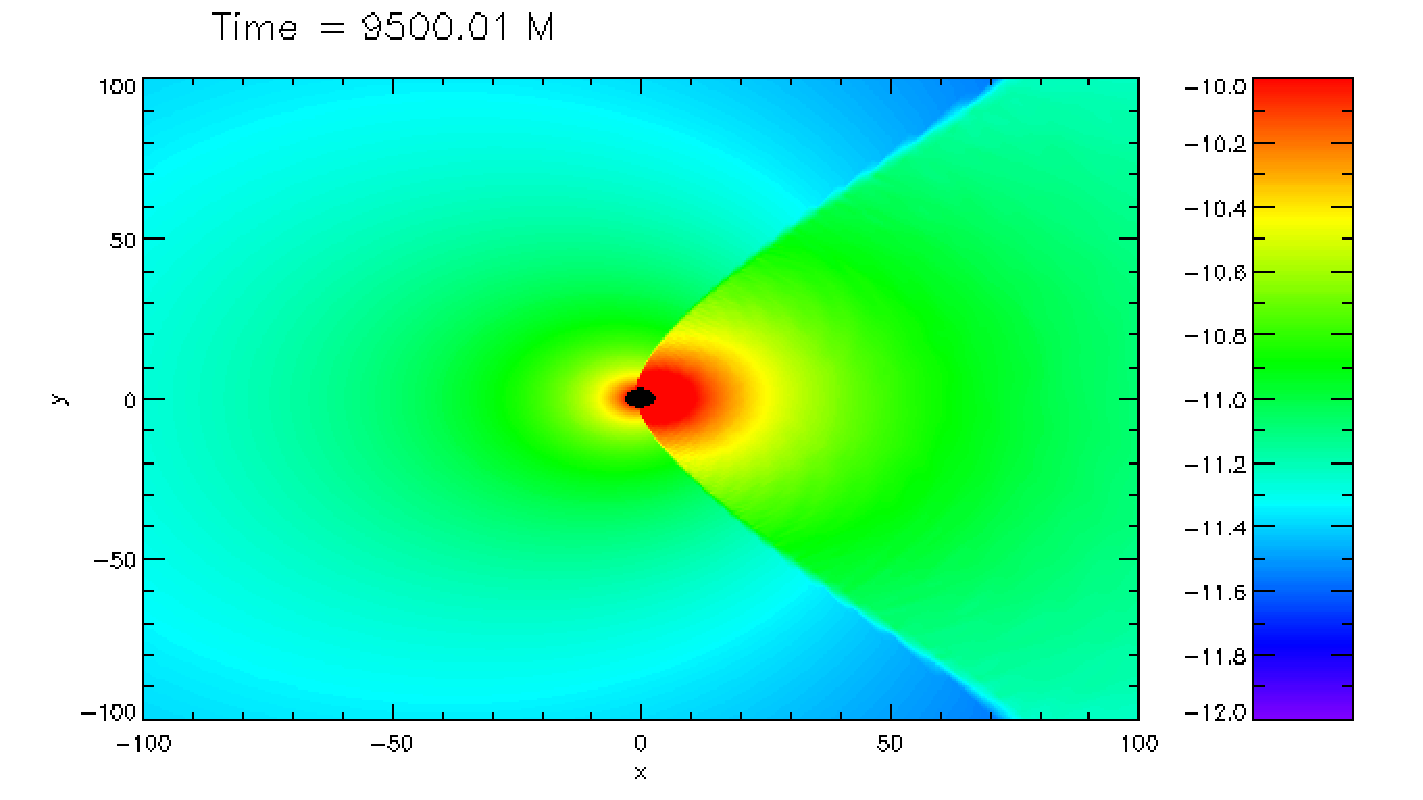}}
{\includegraphics[angle=0,width=8.0cm,height=7.3cm]{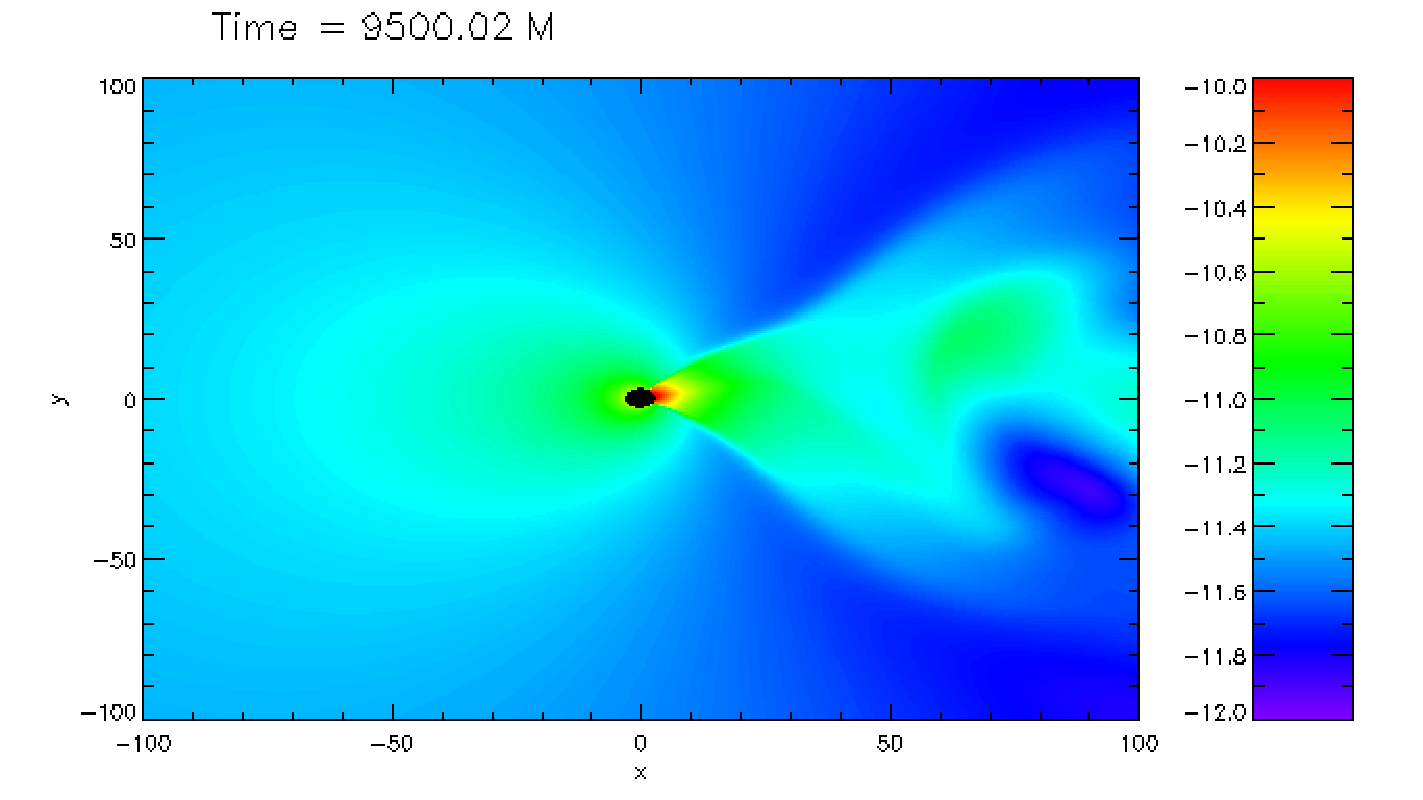}}
{\includegraphics[angle=0,width=8.0cm,height=7.3cm]{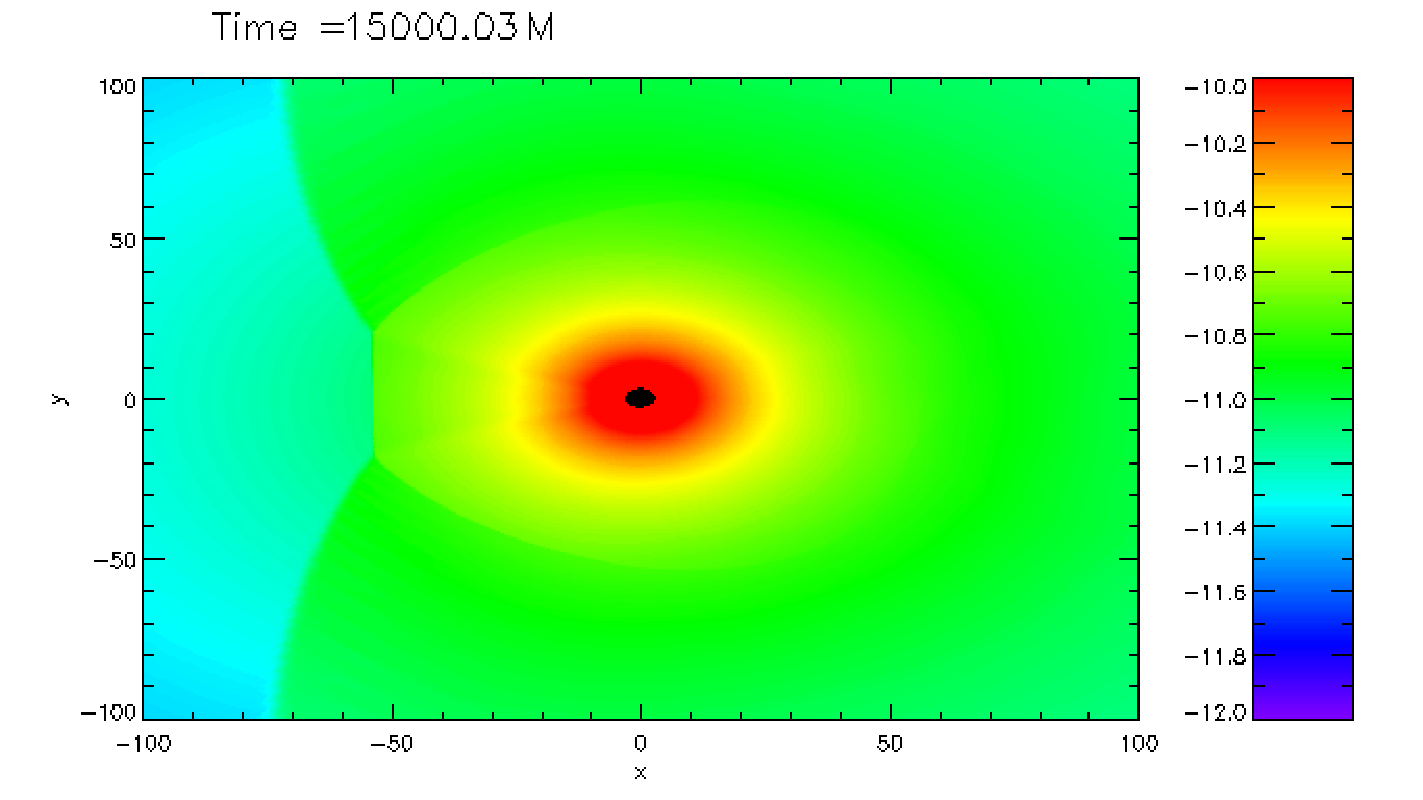}}
{\includegraphics[angle=0,width=8.0cm,height=7.3cm]{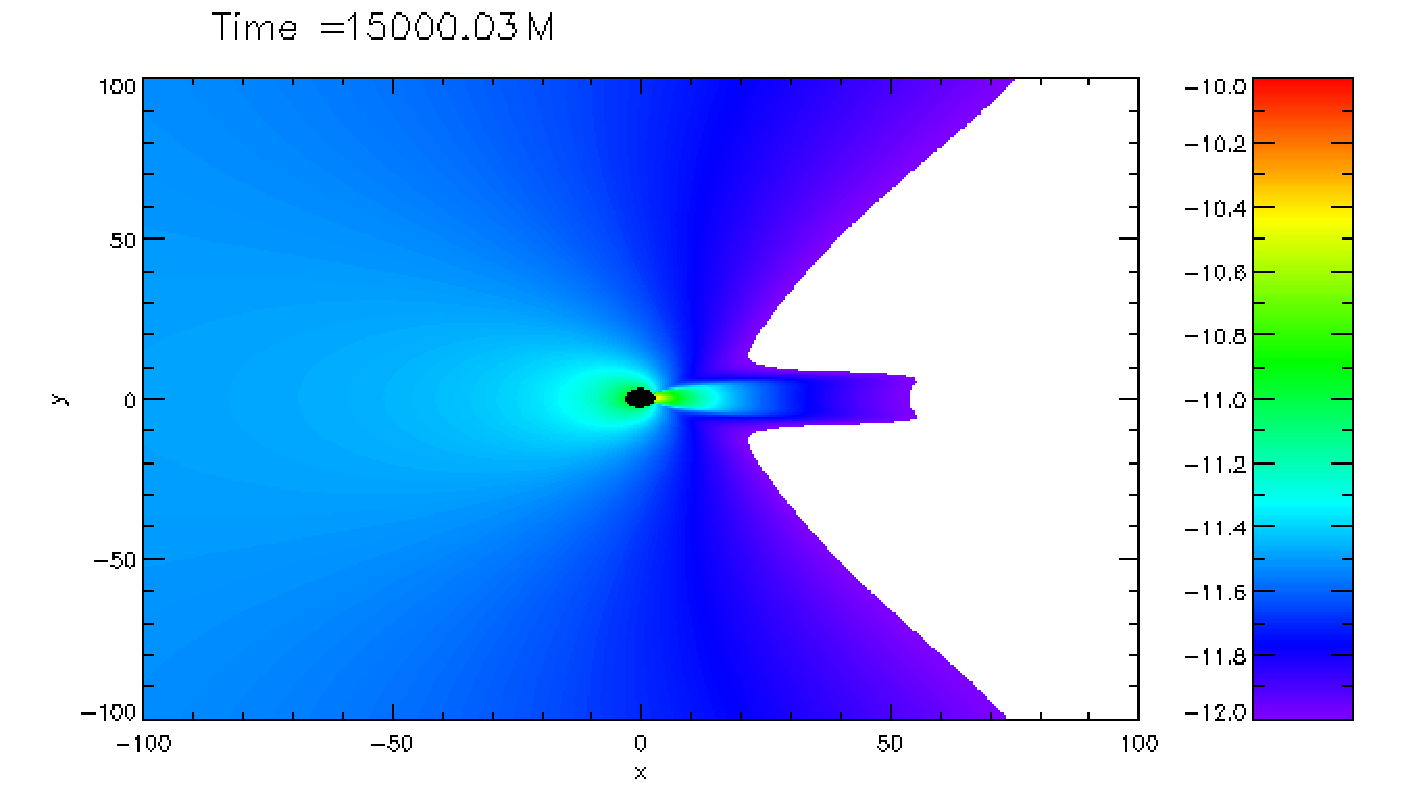}}
\caption{Rest-mass density in cgs units on a logarithmic scale for
  model $\mathtt{V09.CS07}$ in a purely hydrodynamical evolution (left
  panels) and in a radiation-hydrodynamics evolution (right
  panels). Different rows refer to different times of the evolution
  and white regions correspond to densities slightly below the
  threshold for the colour code at around $10^{-12}\,{\rm
    g/cm}^{-3}$. Note that the presence of a radiation field reduces
the rest-mass density  considerably  near the black hole, suppressing
  the accretion rate.}
\label{fig:hydro_vs_rad1}
\end{figure*}

\begin{figure*}
\centering
{\includegraphics[angle=0,width=8.5cm,height=8.0cm]{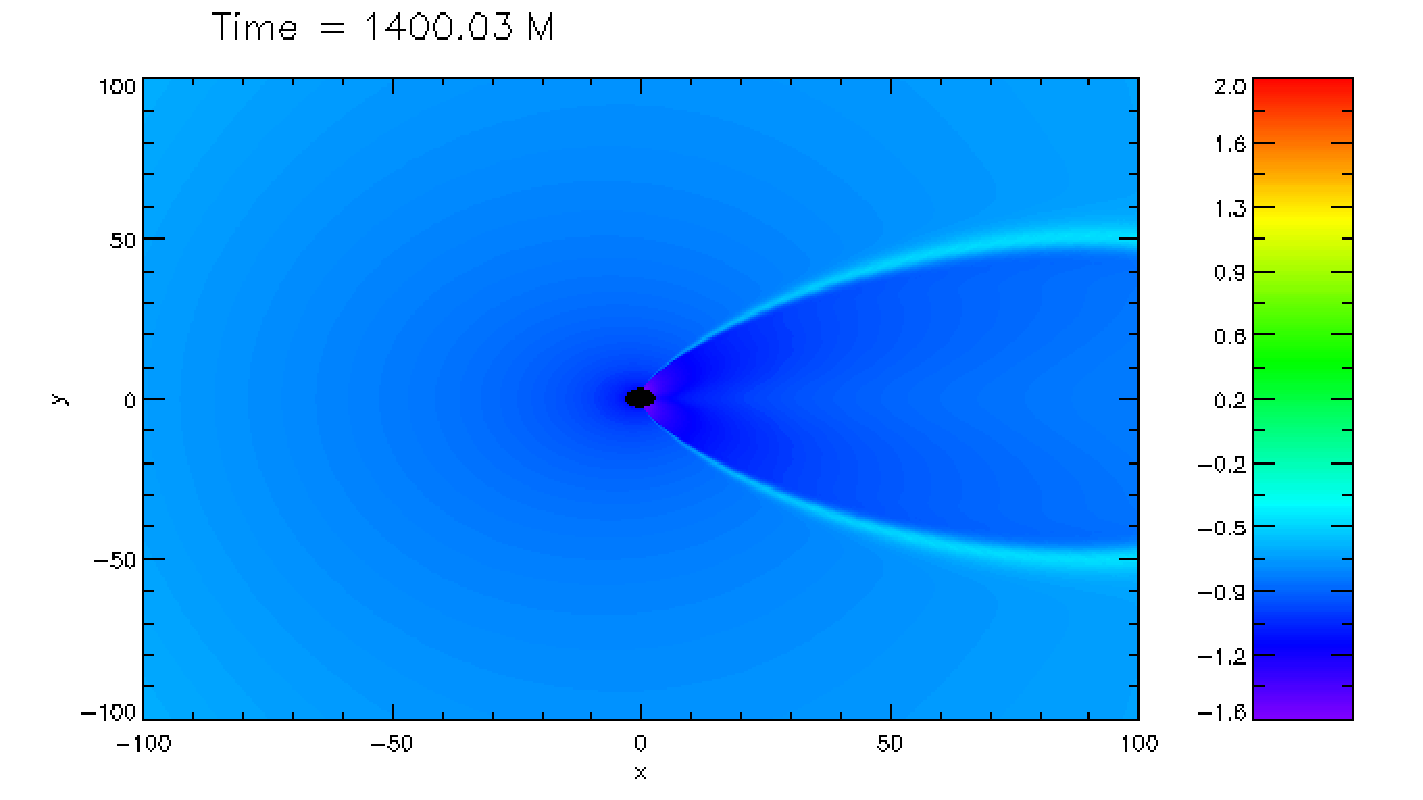}}
\hskip 0.5cm
{\includegraphics[angle=0,width=8.5cm,height=8.0cm]{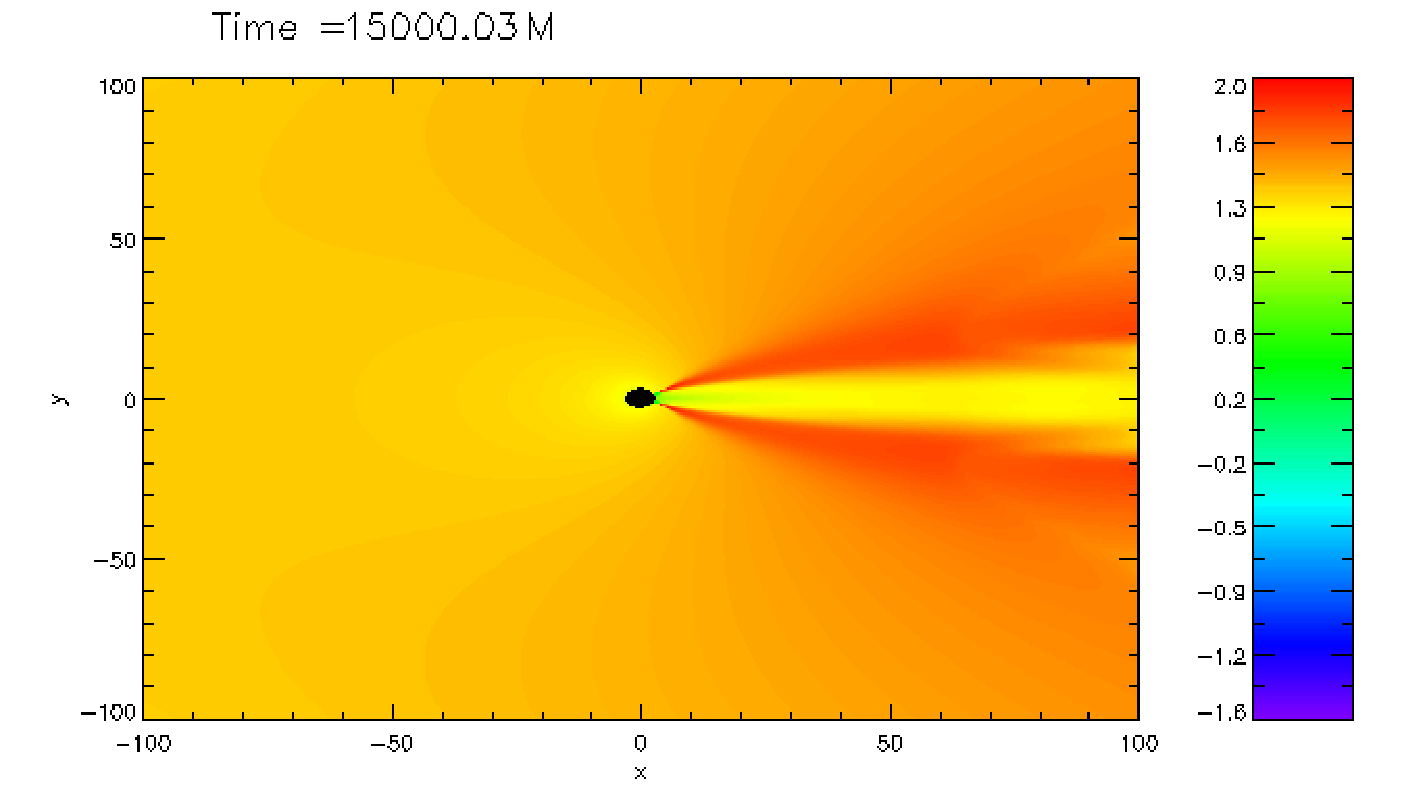}}
\caption{\textit{Left Panel:} logarithm of the ratio of radiation
  pressure over gas pressure for the model $\mathtt{V09.CS07}$ at
  early times.  \textit{Right Panel:} the same as the right panel but
  at later times, when stationarity had been reached.  }
\label{fig:prad_over_p}
\end{figure*}

\begin{figure*}
\centering
{\includegraphics[angle=0,width=8.0cm,height=7.3cm]{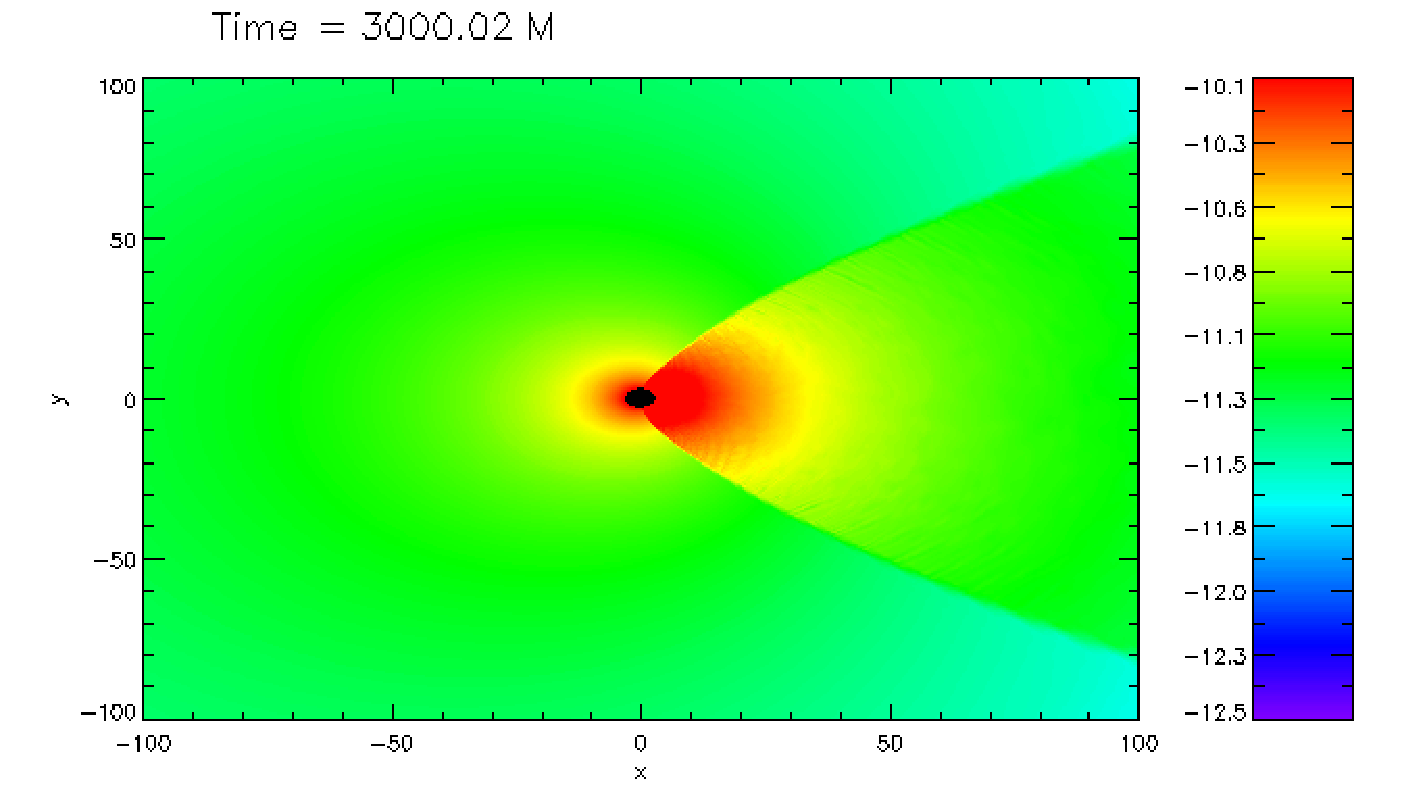}}
{\includegraphics[angle=0,width=8.0cm,height=7.3cm]{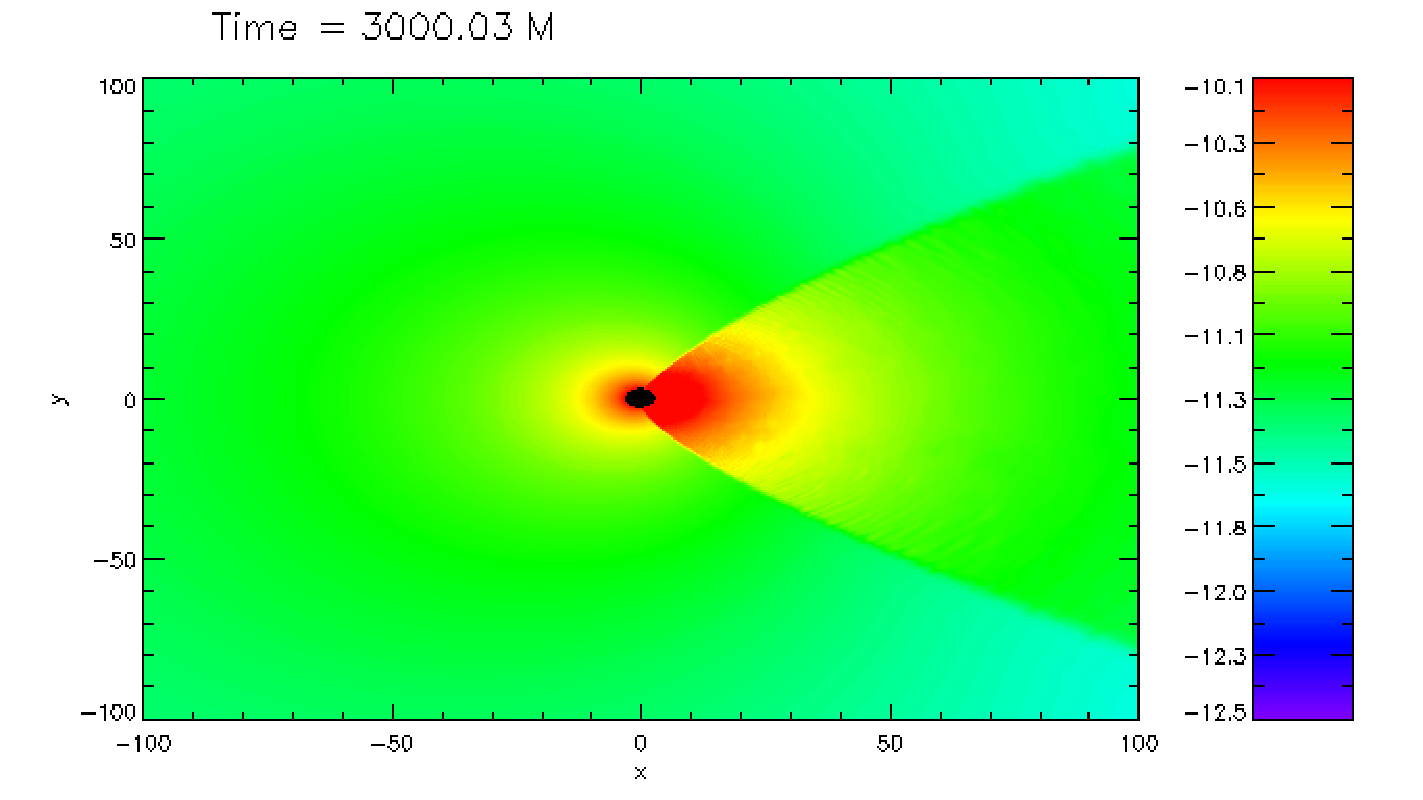}}
{\includegraphics[angle=0,width=8.0cm,height=7.3cm]{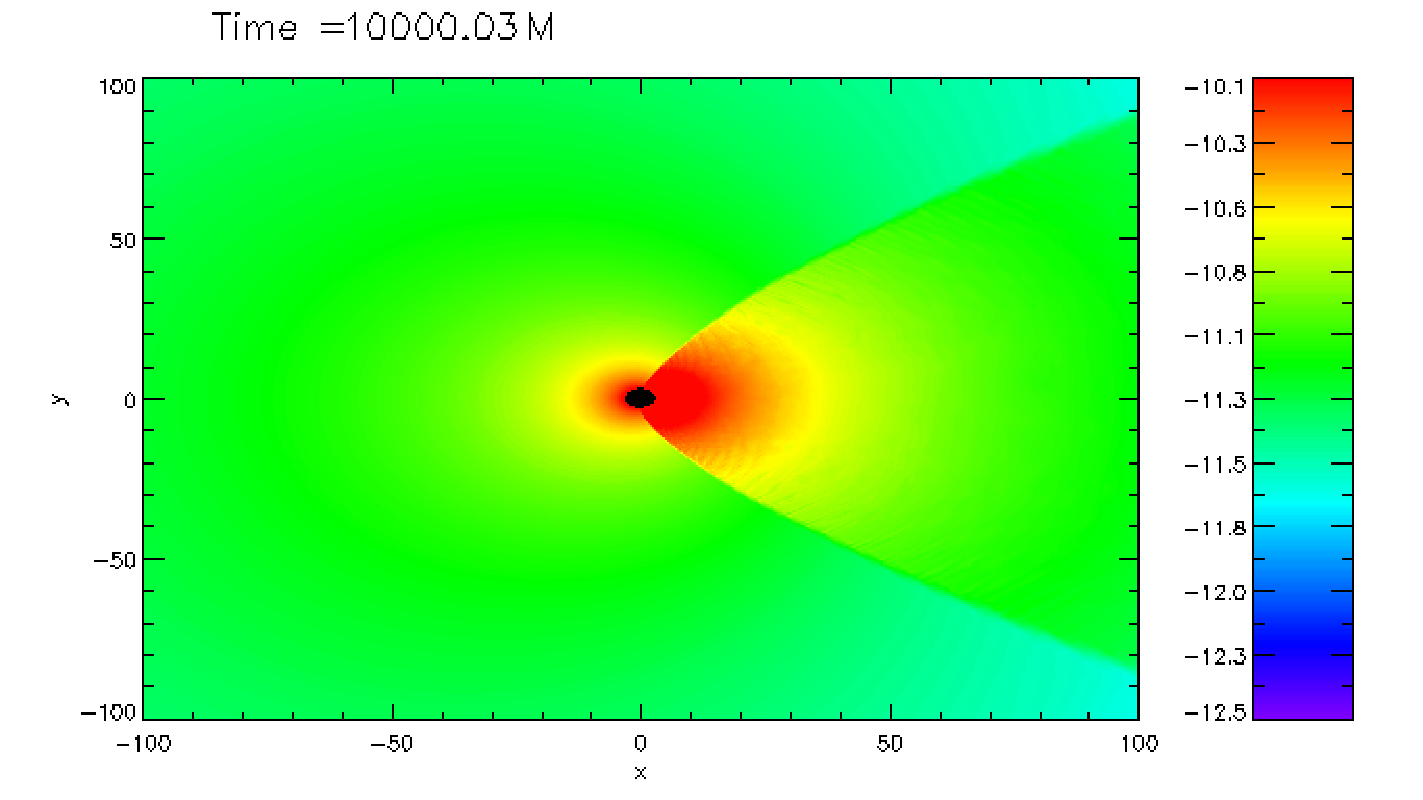}}
{\includegraphics[angle=0,width=8.0cm,height=7.3cm]{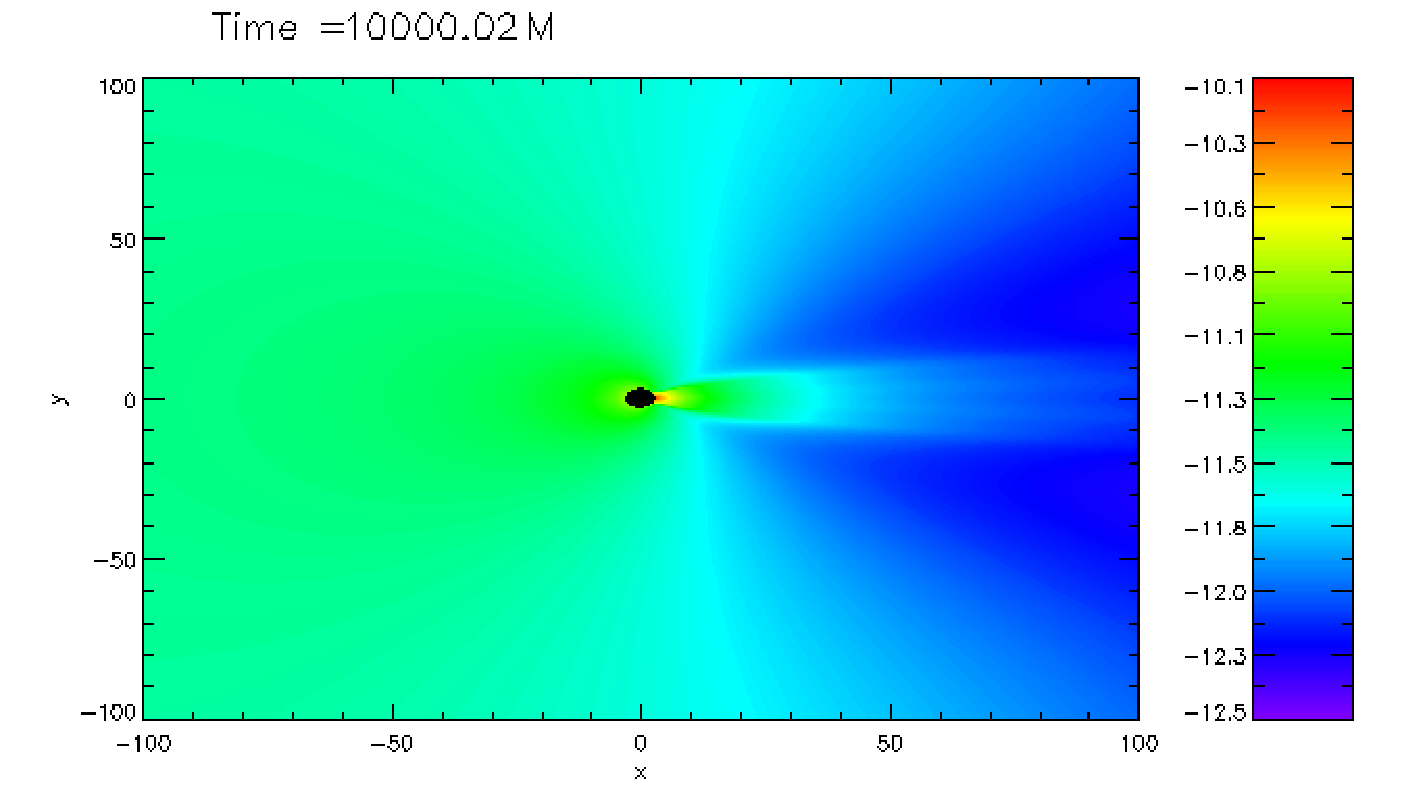}}
{\includegraphics[angle=0,width=8.0cm,height=7.3cm]{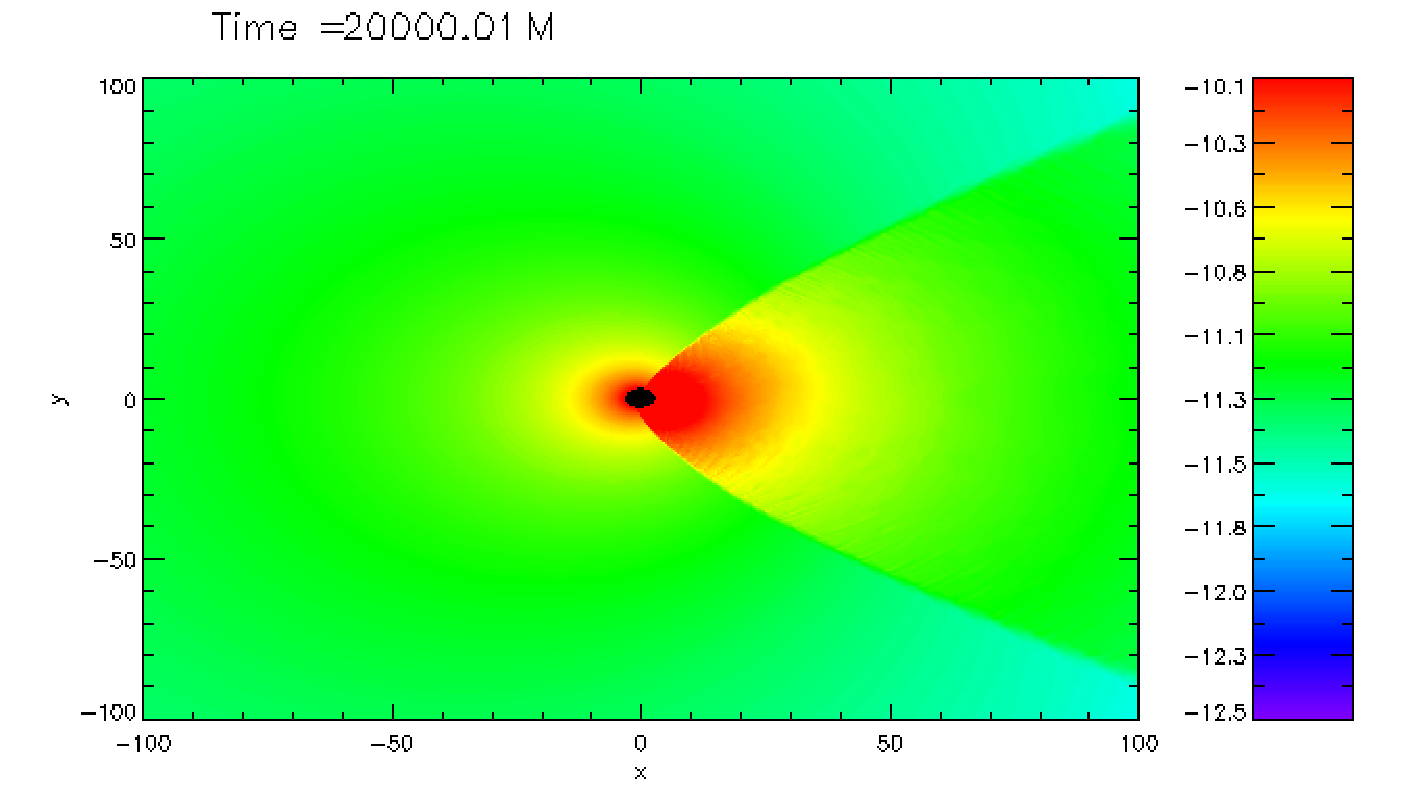}}
{\includegraphics[angle=0,width=8.0cm,height=7.3cm]{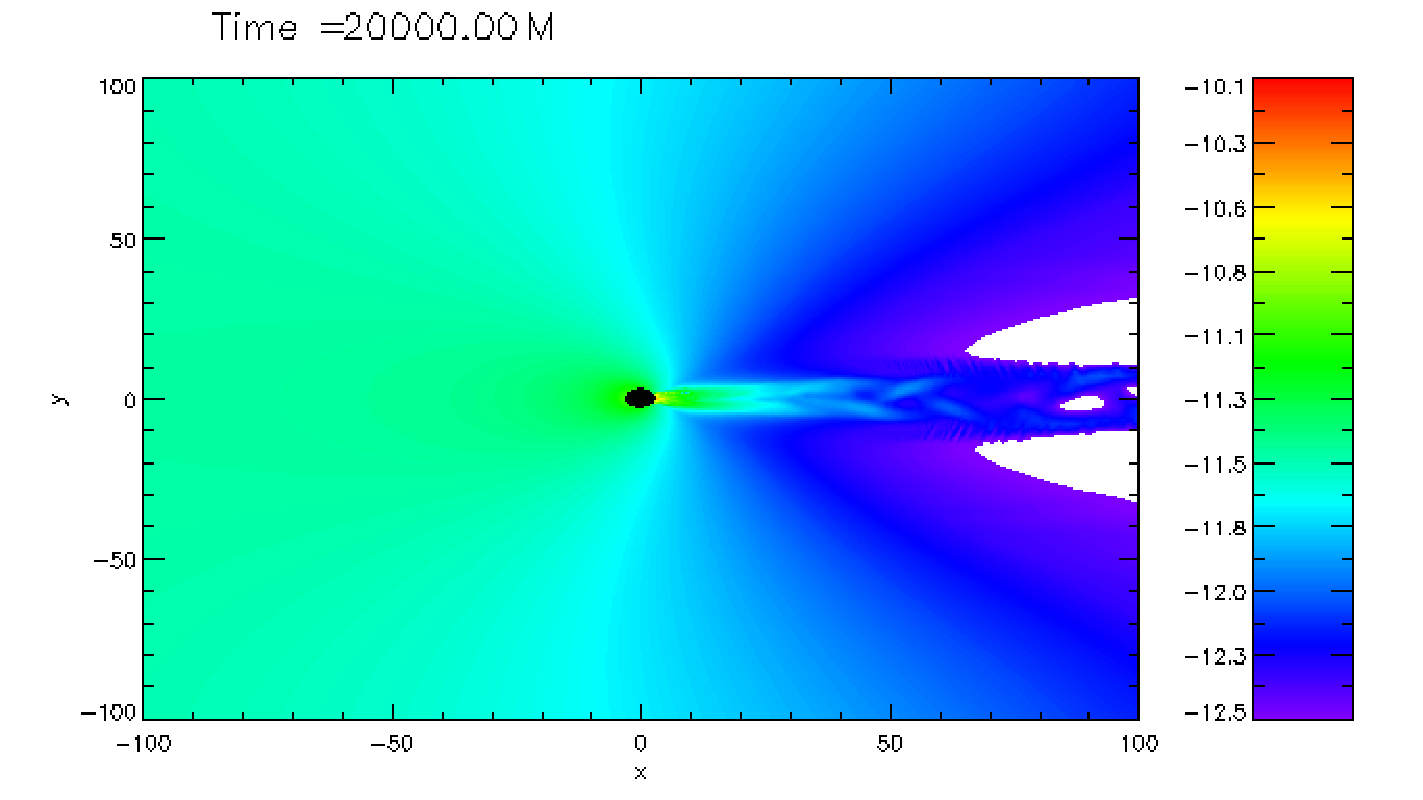}}
\caption{Rest-mass density in cgs units on a logarithmic scale for
  model $\mathtt{V10.CS07}$ in a purely hydrodynamical evolution (left
  panels) and in a radiation-hydrodynamics evolution (right
  panels). Different rows refer to different times of the evolution
  and white regions correspond to densities slightly below the
  threshold for the colour code at around $10^{-12}\,{\rm
    g/cm}^{-3}$. Note that the presence of a radiation field reduces
 the rest-mass density considerably  near the black hole, suppressing
  the accretion rate.}
\label{fig:hydro_vs_rad2}
\end{figure*}

\subsection{Results}

Before entering into the details of our results, it is useful to
briefly review the main features of the relativistic Bondi-Hoyle
accretion as investigated through purely hydrodynamical simulations
by~\citet{Petrich89, Font98a, Font1998g}
and~\citet{Font1999b}. Overall, these studies have highlighted that
when a homogeneous flow of matter moves non-radially towards a compact
object, a shock wave will form close to the accretor. Depending on the
adiabatic index and on the asymptotic Mach number\footnote{We recall
  that the relativistic Mach number is defined as \hbox{${\cal
      M}=\Gamma v/({\rm c}_s\Gamma_s)$}, where $\Gamma$ and $\Gamma_s$
  are the Lorentz factors of the flow and of the sound speed,
  respectively. } ${\cal M}_{\infty}$, the shock can either come very
close to the accretor or be at a certain distance from it [see, for
  instance,~\citet{Foglizzo2005}]. In general, for any given value of
the adiabatic index, there is a minimum asymptotic Mach number above
which a shock wave of conic shape, \ie a \textit{``shock cone''},
forms downstream of the accretor. On the other hand, asymptotic Mach
numbers below the critical value produce a shock wave that, initially
formed in the downstream region, opens progressively and reverses in
the upstream region as a bow shock. More recently, two different
studies have shed additional light on the physics of relativistic
Bondi-Hoyle accretion flows. In the first one,~\citet{Donmez2010},
reported the occurrence of the so called \textit{flip-flop}
instability of the shock cone in the relativistic regime and have also
shown that quasi-periodic oscillations of sonic nature are produced in
the shock cone. In the second one,~\citet{Penner2011} investigated the
effects of a uniform magnetic field, finding that it produces an
increase in the cone opening angle and in the mass accretion rate.

\subsubsection{Classical Bondi-Hoyle accretion}
\label{Unperturbed Bondi-Hoyle flow}

\begin{figure}
\vspace*{-3.5cm}
 {\includegraphics[angle=0,width=7.5cm,height=7.5cm]{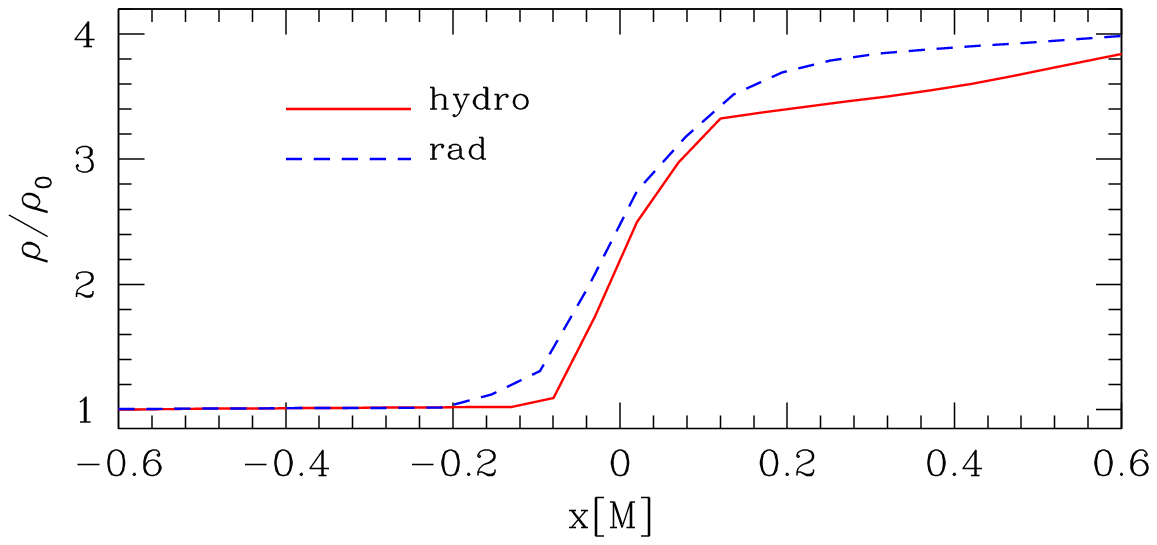}}
\caption{Comparison of the rest-mass density jump across the shock
  front at time $t=9500$ for models $\mathtt{V09.CS07}$ in a purely
  hydrodynamics evolution (red solid line) and in a radiation
  hydrodynamics evolution (blue dashed line). The location of the
  shock was re-normalized to lie at $x=0$ and is displaced by $\Delta
  x\,\sim\,0.2M$ between the two runs.}
\label{fig:shock_jumps}
\end{figure}

We start our analysis by considering the extent to which a radiation
field affects the dynamics of the classical Bondi-Hoyle flow,
comparing the dynamics for very similar physical conditions.
The initial models, which are the first seven reported in
Table~\ref{table:Initial Models}, have very high temperatures and,
consequently, high thermal conductivities\footnote{The present version
  of the code does not allow to handle stiff source terms that arise
  in the radiation-hydrodynamics equations when the conductivity is
  small~\citep{Szoke2006}. Work is in progress to cope with this
  difficulty.}.
As mentioned above, for any given
value of the adiabatic index, there is a critical asymptotic Mach
number ${\cal M}_{\infty,c}$, usually close to unity, above which a
shock cone forms in the downstream region and below which the shock
cone reverses in the upstream region. Our simulations indicate that,
for values of the Mach number close to the critical one, the radiation
effects on the dynamics are most evident. This is shown in
Fig.~\ref{fig:hydro_vs_rad1} for model $\mathtt{V09.CS07}$, where we
have reported the distribution of the rest-mass density at three
different times in a purely hydrodynamical evolution (left panels) and
in a radiation-hydrodynamic evolution (right panels).  This model, in
particular, provides an example in which the radiation field prevents
the reversal of the shock cone from the downstream region into the
upstream region, which instead takes place in the purely
hydrodynamical evolution. Since the dynamics of $\mathtt{V09.CS07}$
becomes radiation-pressure dominated around $t\sim 5000 \, M$, the
explanation of this effect is simple: In such conditions the effective
adiabatic index of the fluid-plus-radiation medium is smaller than
that of the fluid alone [see Eq.~(70.22) of~\citet{Mihalas84}]
\be
\gamma_{\rm{eff}}=\frac{5/2+20q+16q^2}{(3/2+12q)(1+q)} \,,
\ee
where $q={\cal P}_{\rm r}/p$. This fact has two important
consequences. The first one, which we will discuss shortly when
commenting Fig.~\ref{fig:shock_jumps}, is to increase the rest-mass
density jumps across shock fronts. The second one, is exactly to
favour the generation of the shock cone downstream of the accretor, as
firstly noticed by~\citet{Ruffert1996c} and later confirmed
by~\citet{Font98a}.

\begin{figure*}
{\includegraphics[angle=0,width=7.5cm,height=7.5cm]{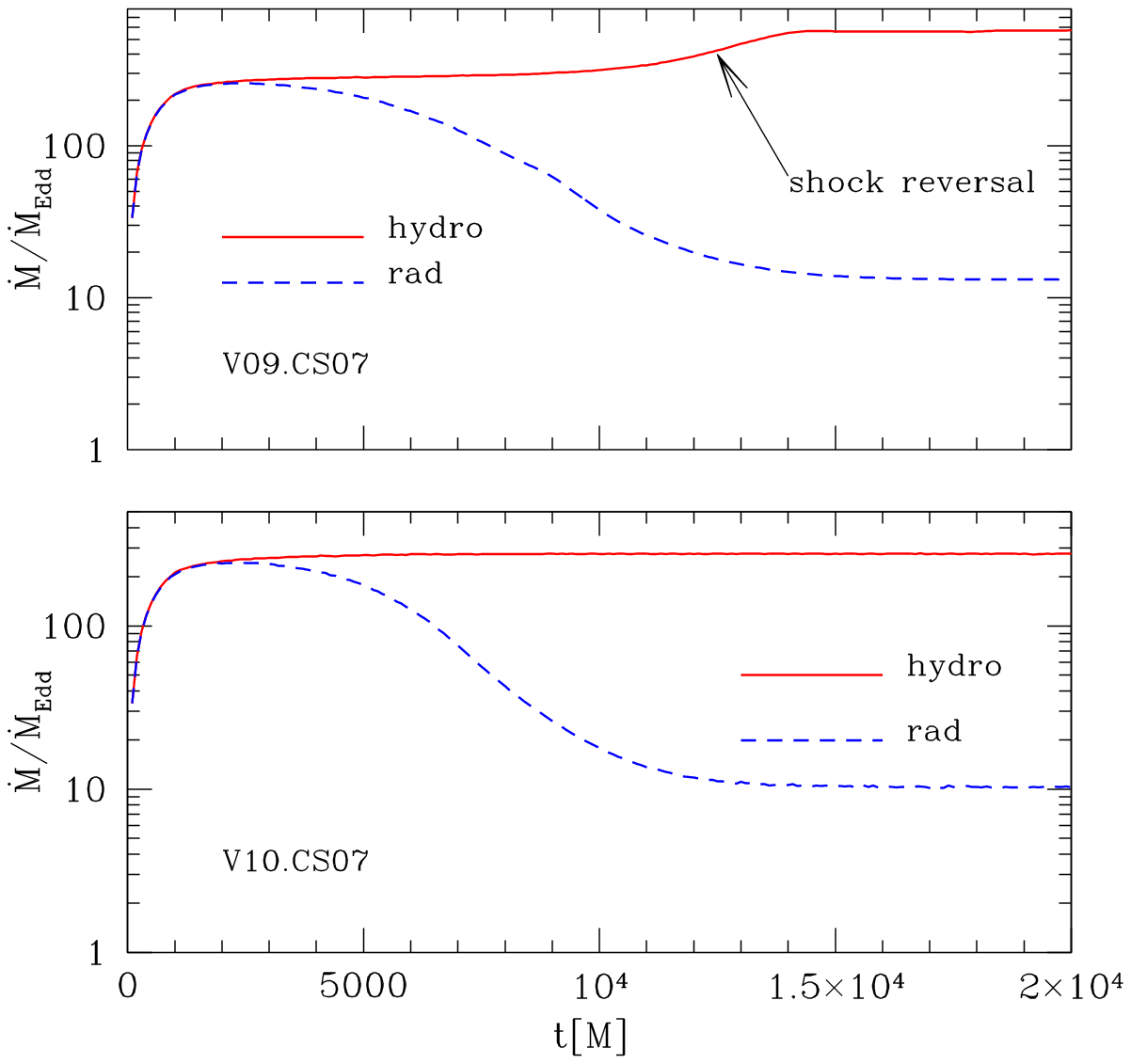}}
\hskip 1.0cm
{\includegraphics[angle=0,width=7.5cm,height=7.5cm]{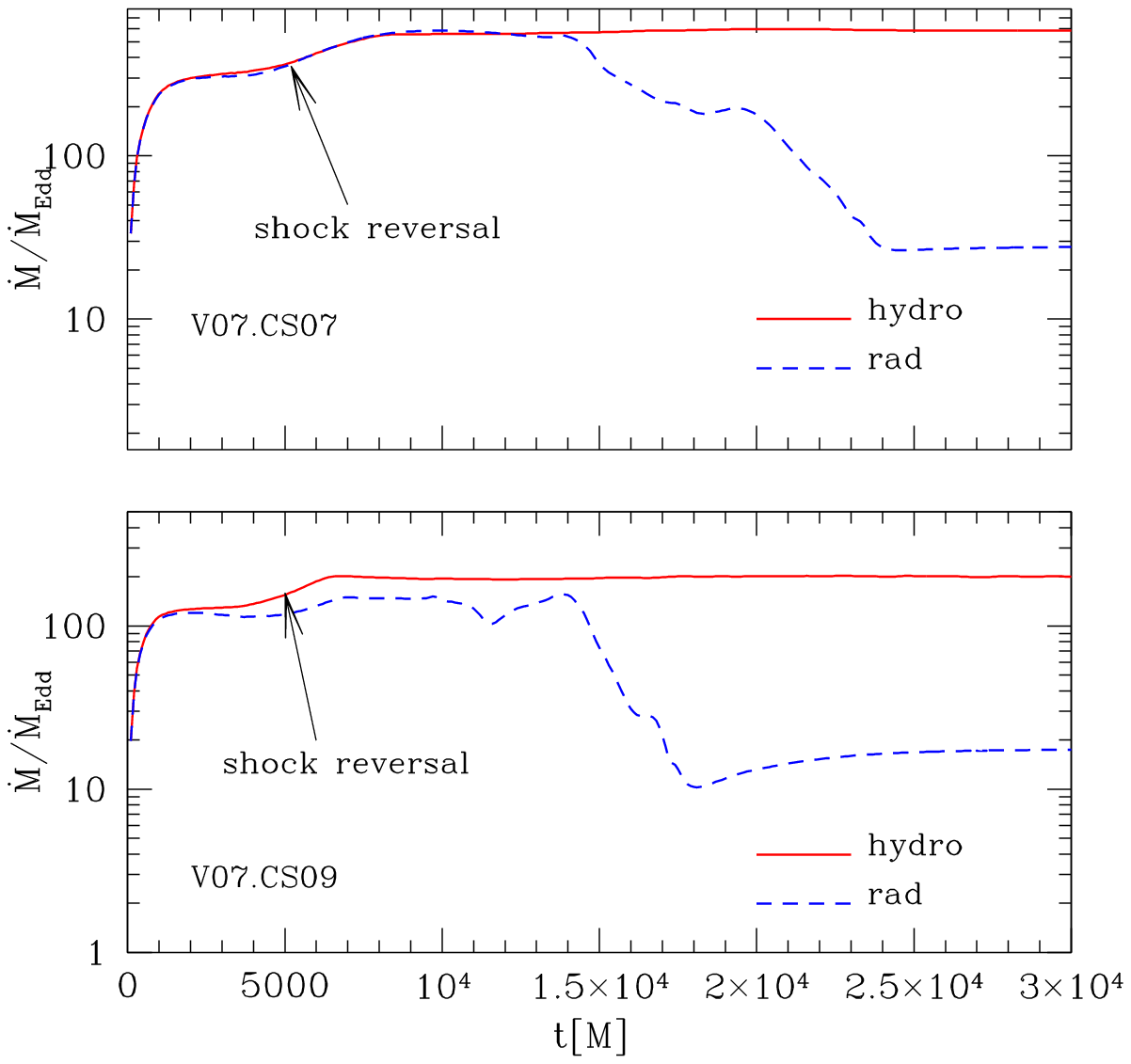}}
\vspace*{-0.0cm}
\caption{Evolution of the mass accretion rate in Eddington units for
  models $\mathtt{V07.CS07}$ and $\mathtt{V07.CS09}$ (left panels) and
  $\mathtt{V09.CS07}$ and $\mathtt{V10.CS07}$ (right panels).  }
\label{fig:accretion_rate}
\end{figure*}

As clearly shown in Fig.~\ref{fig:hydro_vs_rad1}, the
radiation-hydrodynamics evolution of model $\mathtt{V09.CS07}$ is
remarkably different from the purely hydrodynamical one, and it can be
divided in the following stages. After the shock cone has fully opened
in the downstream region (top right panel of
Fig.~\ref{fig:hydro_vs_rad1}), the flow becomes radiation-pressure
dominated, making the shock cone oscillate from one side of the
accretor to the other, in a way that resembles the flip-flop
instability already encountered in relativistic Bondi-Hoyle flows by
\citet{Donmez2010}. This transient behaviour, captured in the
right-middle panel of Fig.~\ref{fig:hydro_vs_rad1}, is accompanied by
an outflow of matter expelled by radiation pressure beyond the
computational grid. After that, the system relaxes to a stationary
configuration characterized by the presence of a shock cone with a
much smaller opening angle than in the hydrodynamical solution, giving
rise to a ``reduced'' shock cone. Also shown in
Fig.~\ref{fig:prad_over_p} is the ratio between the radiation pressure
and the fluid pressure, reported in the left panel for an early and
fluid-pressure dominated stage of the evolution, and in the right
panel for a late and radiation-pressure dominated one.

A very similar behaviour to the one discussed so far is shown in
Fig.~\ref{fig:hydro_vs_rad2} for model $\mathtt{V10.CS07}$, the
initial Mach number of which is only slightly larger than model
$\mathtt{V09.CS07}$. However, in this case the higher fluid velocity
causes supercritical behaviour both in the hydrodynamical and the radiation-hydrodynamical
evolution so that the shock cone remains in the downstream
region. The close similarity between the dynamics of models
$\mathtt{V09.CS07}$ and $\mathtt{V10.CS07}$ in the presence of
 the radiation field is also testified by the asymptotic mass accretion
rate, which is $\dot M\simeq 13.14 \dot M_{\rm Edd}$ for model
$\mathtt{V09.CS07}$ and $\dot M\simeq 10.24 \dot M_{\rm Edd}$ for
$\mathtt{V10.CS07}$.

An information complementary to that of Fig.~\ref{fig:hydro_vs_rad1} -
\ref{fig:hydro_vs_rad2} is provided by Fig.~\ref{fig:accretion_rate},
which shows the evolution of the mass accretion rate for a few
selected models. For each of these models both the purely
hydrodynamical evolution (red solid lines) and the
radiation-hydrodynamical one (blue dashed lines) are considered. A few
comments are worth making about this figure. The first one is that,
once stationarity is reached, the mass accretion rates of the
radiation-hydrodynamics models are significantly smaller than those of
the corresponding hydrodynamics models. This result was of course
expected, because of the obstructive effect of the radiation
pressure. The second comment is that the reversal of the shock cone in
the hydrodynamics models $\mathtt{V09.CS07}$, $\mathtt{V07.CS07}$,
$\mathtt{V07.CS09}$ and in the radiation-hydrodynamics models
$\mathtt{V07.CS07}$ and $\mathtt{V07.CS09}$ leads to an increase of
$\dot M$, as highlighted by the arrows.  For the hydrodynamical
version of model $\mathtt{V09.CS07}$, for instance, this increase
starts at $t\sim \, 12000 \, M$, as reported in the top-left panel of
Fig.~\ref{fig:accretion_rate}. Finally, we find that all models
accrete at super-Eddington rates even when a radiation field is
present. This is not surprising, since the Eddington limit holds
strictly only in spherical symmetry, which is not fulfilled in
wind-like accretion. Moreover, it should be remarked that the
classical Eddington limit is computed in a framework where only the
electron Thomson cross section contributes to the radiation pressure.

\begin{figure*}
\centering
{\includegraphics[angle=0,width=8.5cm,height=8.0cm]{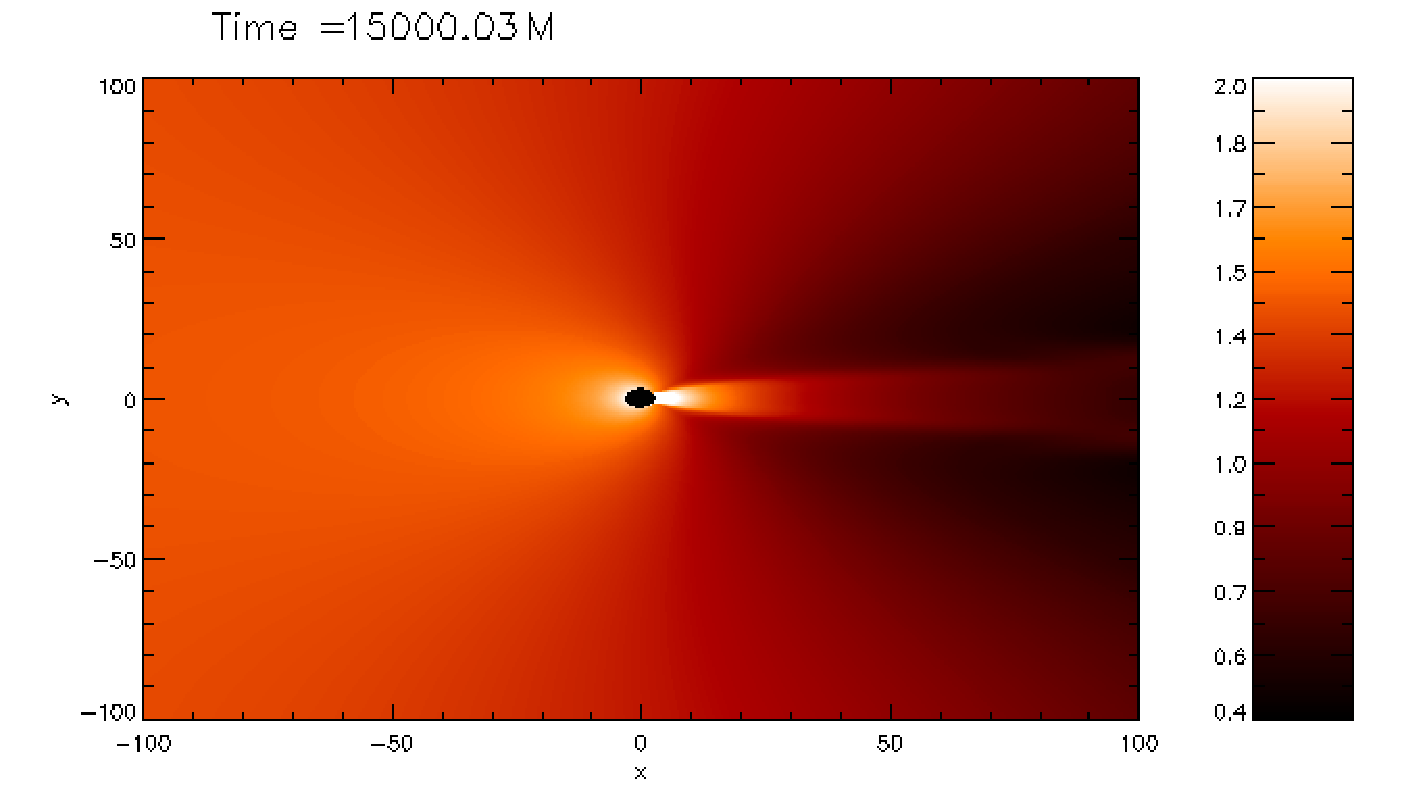}}
\hskip 0.5cm
{\includegraphics[angle=0,width=8.5cm,height=8.0cm]{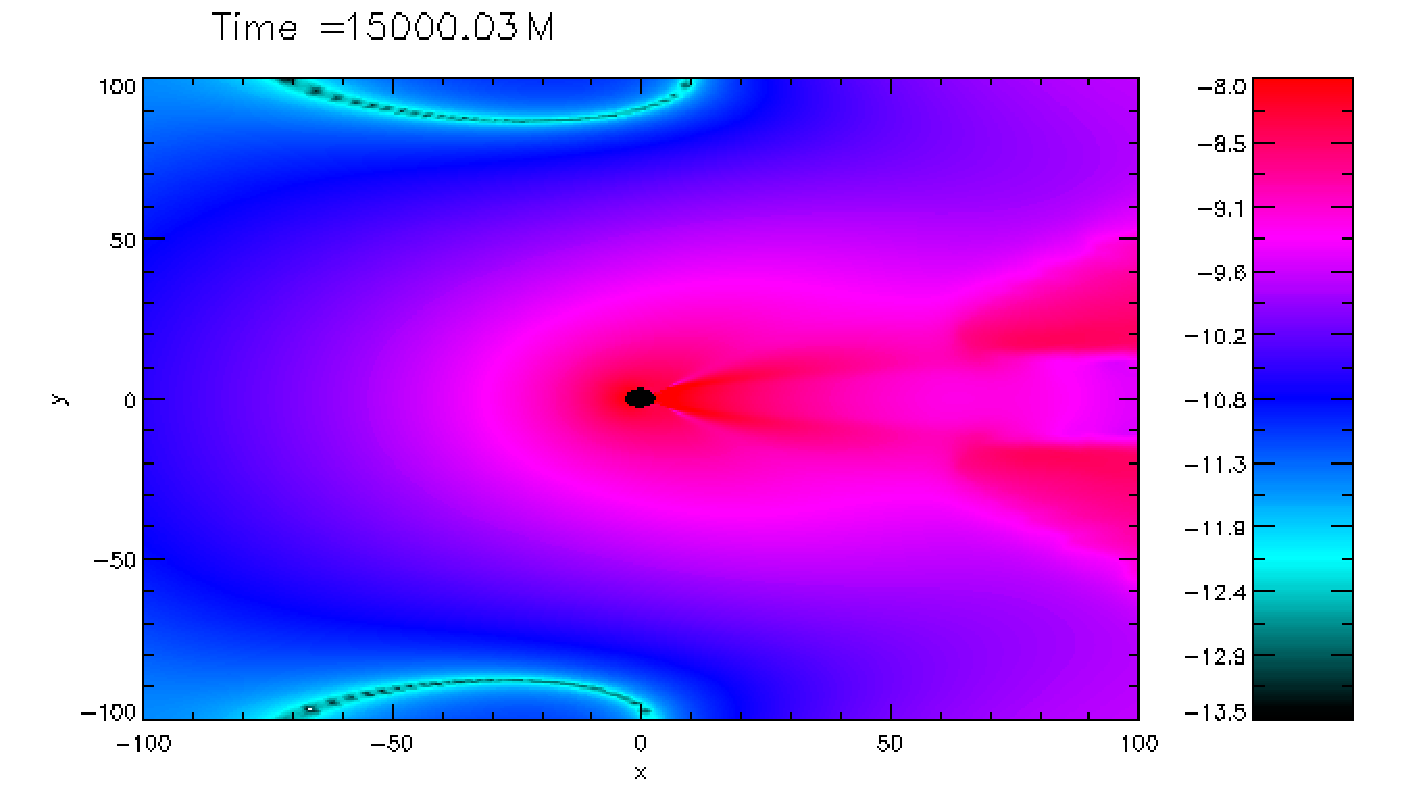}}
\caption{\textit{Left Panel:} logarithm of the optical thickness for
  the model $\mathtt{V09.CS07}$ once stationarity is
  reached. \textit{Right Panel:} Logarithm of the modulus of the
  radiative flux (in geometrized units) in the same model at the same
  time. }
\label{fig:opt_flux}
\end{figure*}

Additional differences between the hydrodynamics and the
radiation-hydrodynamics evolutions emerge after comparing the jumps
experienced by the rest-mass density across a shock wave in a
representative model. Such a comparison is reported in
Fig.~\ref{fig:shock_jumps} for model $\mathtt{V09.CS07}$, showing the
variation of the rest-mass density across the shock that is produced
at time $t=5000 M$ and visible in the two top panels of
Fig.~\ref{fig:hydro_vs_rad1}. The two curves have been obtained after
slicing the rest-mass density along an ``$x$-direction'' perpendicular
to the shock front, and sliding the two profiles so that the shock is
located at the same $x=0$ position for both the hydrodynamical and the
radiation-hydrodynamical evolution. Negative and positive values of
the $x$-coordinate refer therefore to the unshocked and to the shocked
region, respectively, while the rest-mass density has been normalized
to the value in the unshocked region.

The first comment about this figure is that the density
jump in the hydrodynamics evolution is slightly smaller
than the value of $4$ predicted by the theoretical
expectation of $\rho_2/\rho_1\sim(\gamma+1)/(\gamma-1)$
valid for an ideal gas EOS. This effect may be due 
to the presence of both numerical diffusion and
tangential velocities along the shock front.
The second comment is 
that the compression
ratio across the shock increases by $3\%$ in the transition from a
hydrodynamical to a radiation-hydrodynamical
evolution. This effect can again be understood 
by regarding the fluid in the radiation--hydrodynamics
evolution as an effective fluid having a \textit{smaller} adiabatic index,
as indeed expected in the radiation-pressure dominated
regime. This result is also in agreement with analytical
investigations by~\citet{Guess1960} and \cite{Mihalas84}
(\textsection 104).

\begin{figure*}
{\includegraphics[angle=0,width=7.5cm,height=7.5cm]{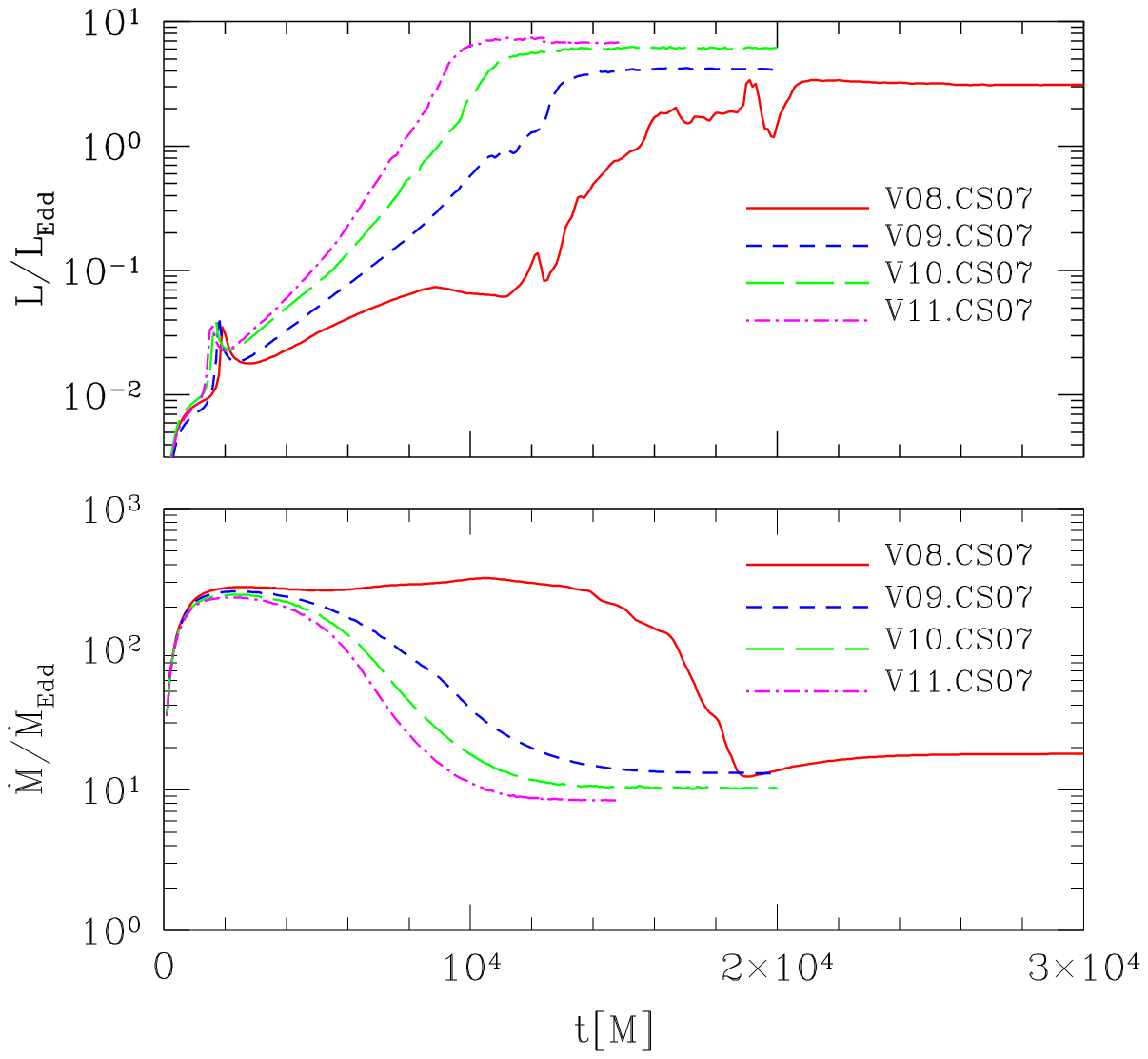}}
\hskip 1.0cm
{\includegraphics[angle=0,width=7.5cm,height=7.5cm]{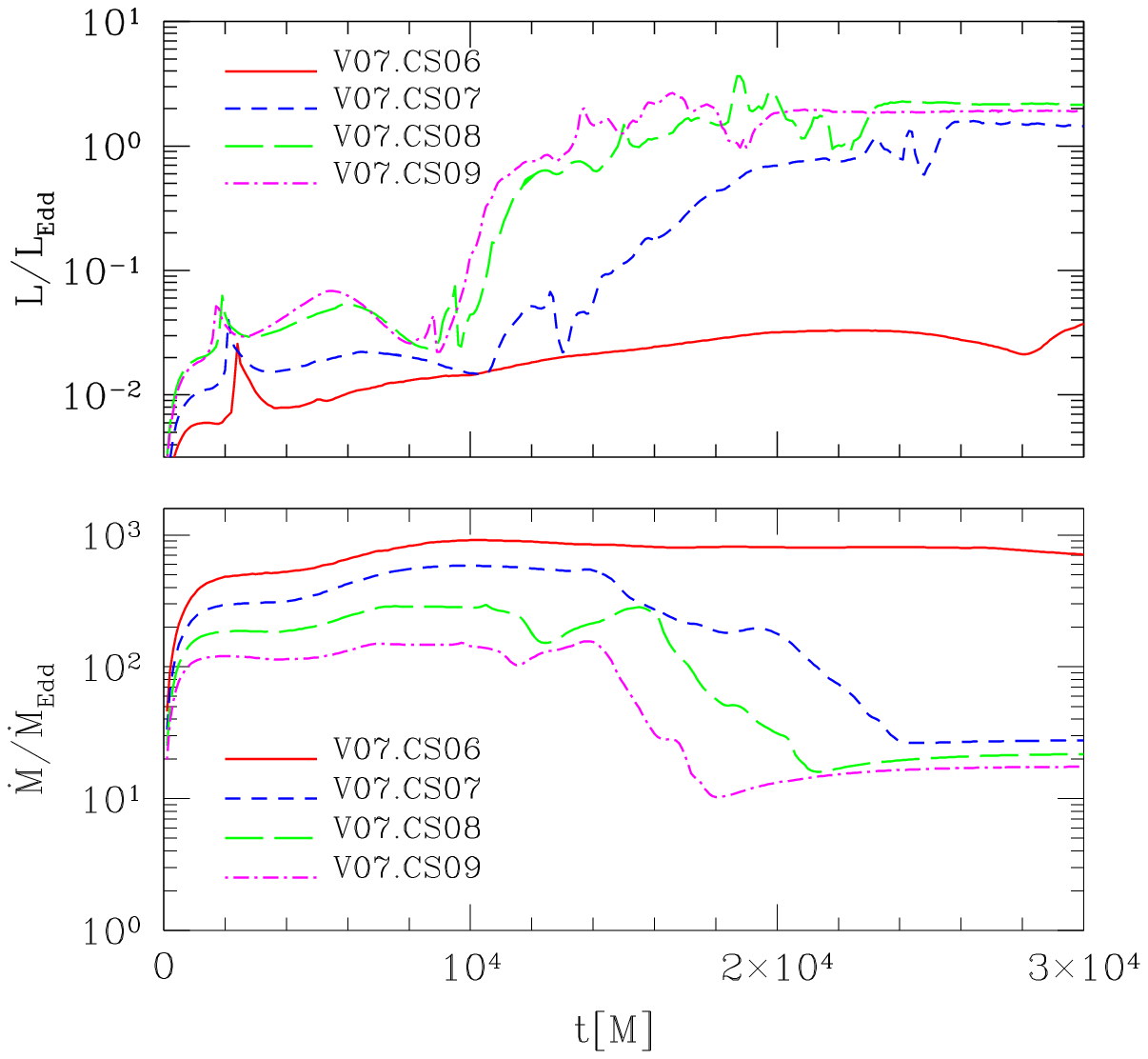}}
\vspace*{-0.0cm}
\caption{Luminosity and mass-accretion rates in Eddington units in
  classical Bondi-Hoyle accretion flows. The left panel collects
  models with different initial velocities but with the same sound
  speed. In contrast, the right panel collects models with the same
  initial velocities but with different values for the sound
  speed. All simulations were evolved until they reached stationarity,
  and up to $t=30000\,M$ at most.}
\label{fig:light_curve}
\end{figure*}

%
\begin{table}
  \caption{Mass accretion rate, luminosity and efficiency
    $\eta_{_{\cal BH}}$ as defined in Eq.~(\ref{eq:defeta}) of
    Bondi-Hoyle accretion in the quasi-stationary regime.
 \label{table:eta}}
\begin{center}
  \begin{tabular}{ccccc}
    \hline \hline
   Model & ${\cal M}_{\infty}$  & $\dot{M}_{\rm acc}/\dot{M}_{\rm {Edd}}$ &
   $L/L_{\rm {Edd}}$ & $\eta_{_{\cal BH}}$ \\
    \hline
$\mathtt{V08.CS07}$ & $1.14$ & $17.9$ & $3.10$ &$0.14$ \\
$\mathtt{V09.CS07}$ & $1.28$ & $13.14$ & $4.11$ &$0.22$ \\
$\mathtt{V10.CS07}$ & $1.42$ & $10.24$ & $6.18$ &$0.36$ \\
$\mathtt{V11.CS07}$ & $1.57$ & $8.32$ & $6.77$ &$0.38$ \\
\hline
$\mathtt{V07.CS07}$ & $1.0$  & $27.64$ & $1.44$ &$0.05$ \\
$\mathtt{V07.CS08}$ & $0.87$ & $21.74$ & $2.14$ &$0.09$ \\
$\mathtt{V07.CS09}$ & $0.77$ & $17.44$  & $1.91$ &$0.10$ \\
\hline
$\mathtt{p.V10.CS07}$ & $1.42$ & $11.58$  & $5.35$   &$0.29$ \\
$\mathtt{p.V11.CS07}$ & $1.57$ & $8.34$  & $7.05$   &$0.40$ \\
$\mathtt{p.V18.CS07}$ & $2.57$ & $4.02$  & $30.5$  &$0.69$ \\
   \hline
    \hline
  \end{tabular}
\end{center}
\end{table}

Before computing the luminosity as described in
Sec.~\ref{subsec:luminosity}, it is important to make sure that the
physical conditions chosen correspond to those required by the code,
namely the presence (and the persistence) of an optically thick
regime. Of course all of the models considered in our simulations and
reported in Table~\ref{table:Initial Models} are in such a physical
regime, with only very limited regions where the optical thickness can
be $\sim {\cal O}(1)$ during the evolution. As a representative
example, the left panel of Fig.~\ref{fig:opt_flux} shows the optical
thickness when the system as relaxed to stationarity, for the same
model $\mathtt{V09.CS07}$ that we have extensively described so far.
The right panel of Fig.~\ref{fig:opt_flux}, on the other hand, shows
the corresponding intensity of the momentum of the radiation
field. After comparing with the right-bottom panel of
Fig.~\ref{fig:hydro_vs_rad1}, it is easy to realize that the
distribution of the radiative fluxes is obviously correlated with the
rest-mass density distribution, but also that a good portion of the
radiative emission is concentrated along the shock fronts of the
reduced shock cone.

The evolution of the emitted luminosity and of the mass-accretion
rates are illustrated in the two panels of
Fig.~\ref{fig:light_curve}. More specifically, the left panel, which
reports models with increasing Mach number but having the same initial
temperature, shows that the luminosity increases with ${\cal
  M}_{\infty}$ and reaches stationary values of a few Eddington
units. On the other hand, the right panel, which reports models with
the same asymptotic velocity but different temperatures, shows that
stationarity is reached on longer timescales and a correlation with
the final luminosity is less robust. In all of the models shown, the
first bump around $t\sim 2000 \, M$ is due to the initial opening of
the shock cone.

By providing the first self-consistent computation of the luminosity
in a Bondi-Hoyle accretion flow, our calculations allow us to derive
the efficiency of the accretion flow $\eta_{_{\cal BH}}$. We remark
that the concept of $\eta_{_{\cal BH}}$ for a Bondi-Hoyle flow, with a
nonzero velocity of the matter at infinity, is not the same as in
standard accretion discs, where the gas flow is supposed to start from
matter at rest at infinity. Thus, we define an effective Bondi-Hoyle
luminosity efficiency $\eta_{_{\cal BH}}$ as
\be 
\eta_{_{\cal BH}}=\frac{L} {\dot{M}_{\rm acc} c^2 + 
\frac{1}{2}\dot{M}_{\infty} v^2_{\infty}}\,,
\label{eq:defeta}
\ee
where the denominator takes into account a kinetic contribution to the
energy flux. We report the values of $\dot{M}_{\rm acc}/\dot{M}_{\rm
  {Edd}}$, $L/L_{\rm {Edd}}$ and $\eta_{_{\cal BH}}$ in
Table~\ref{table:eta} for those models presenting a quasi-stationary
accretion pattern.  From the data reported in the Table it is possible
to deduce the existence of two different regimes in a radiative
Bondi-Hoyle accretion flow. A first regime, corresponding to ${\cal
  M}_{\infty}\lesssim 1$, where the luminosity is dominated by the
accretion-powered luminosity and thus proportional to $\dot{M}$. A
second regime, corresponding to ${\cal M}_{\infty}\gtrsim 1$, where
the luminosity is instead dominated by the emission at the shock
front. In particular, by comparing the first four models that have the
same initial asymptotic sound speed, we note that, as the asymptotic
Mach number is increased, the accretion rates decrease. This effect is
due to the reduced opening angle of the shock cone. The corresponding
luminosity, on the other hand, increases, because of the enhanced
dissipation at the shock front.

As a final remark we note that, as already discussed in
Sec.~\ref{subsec:luminosity}, the luminosities we have reported here
can only provide lower limits on the energy efficiency $\eta_{_{\cal
    BH}}$. We have in fact neglected not only viscous dissipative
processes from the accretion flow, but also any nonthermal emission,
such as inverse Compton or synchrotron radiation, which could arise
from a corona developing near the black hole.

\begin{figure*}
\centering
{\includegraphics[angle=0,width=8.5cm,height=8.0cm]{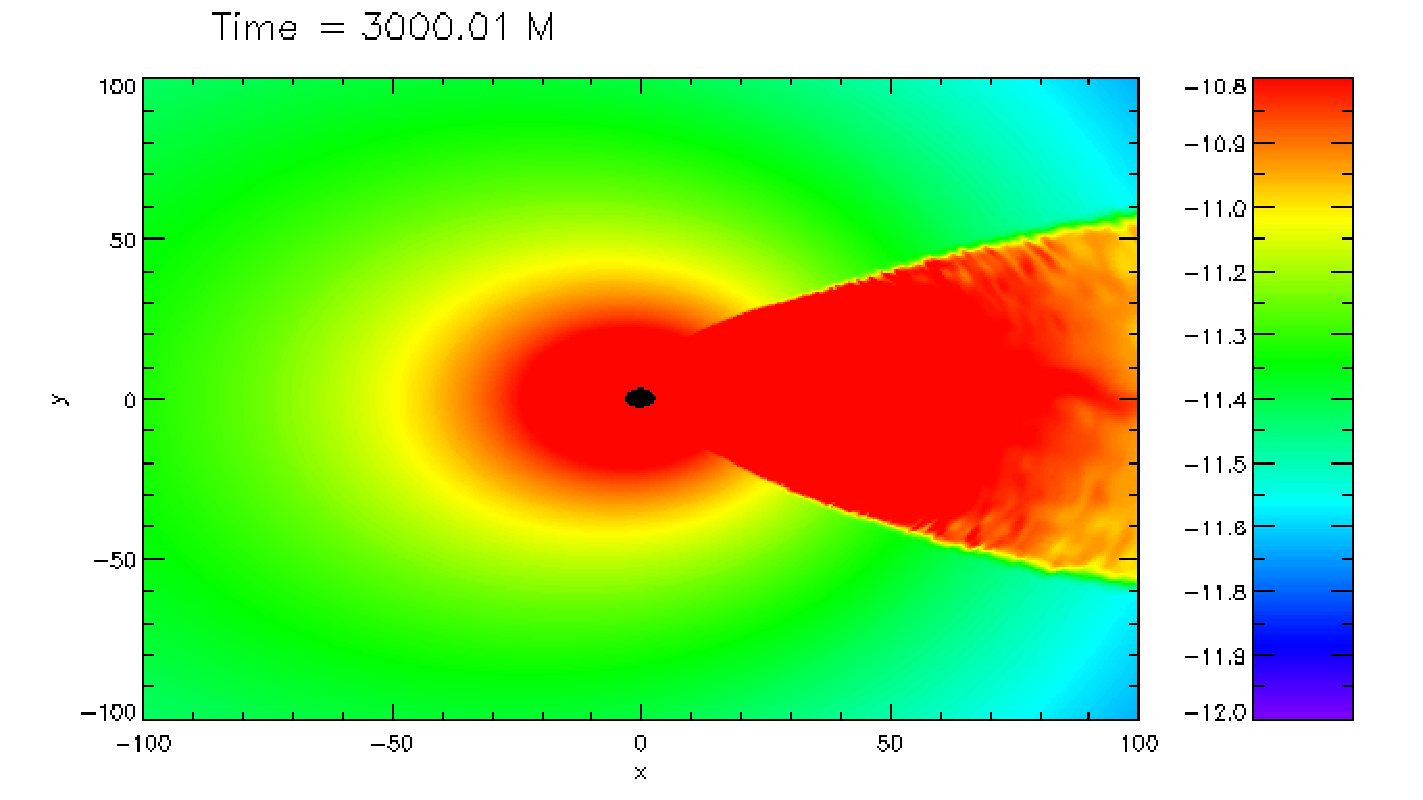}}
{\includegraphics[angle=0,width=8.5cm,height=8.0cm]{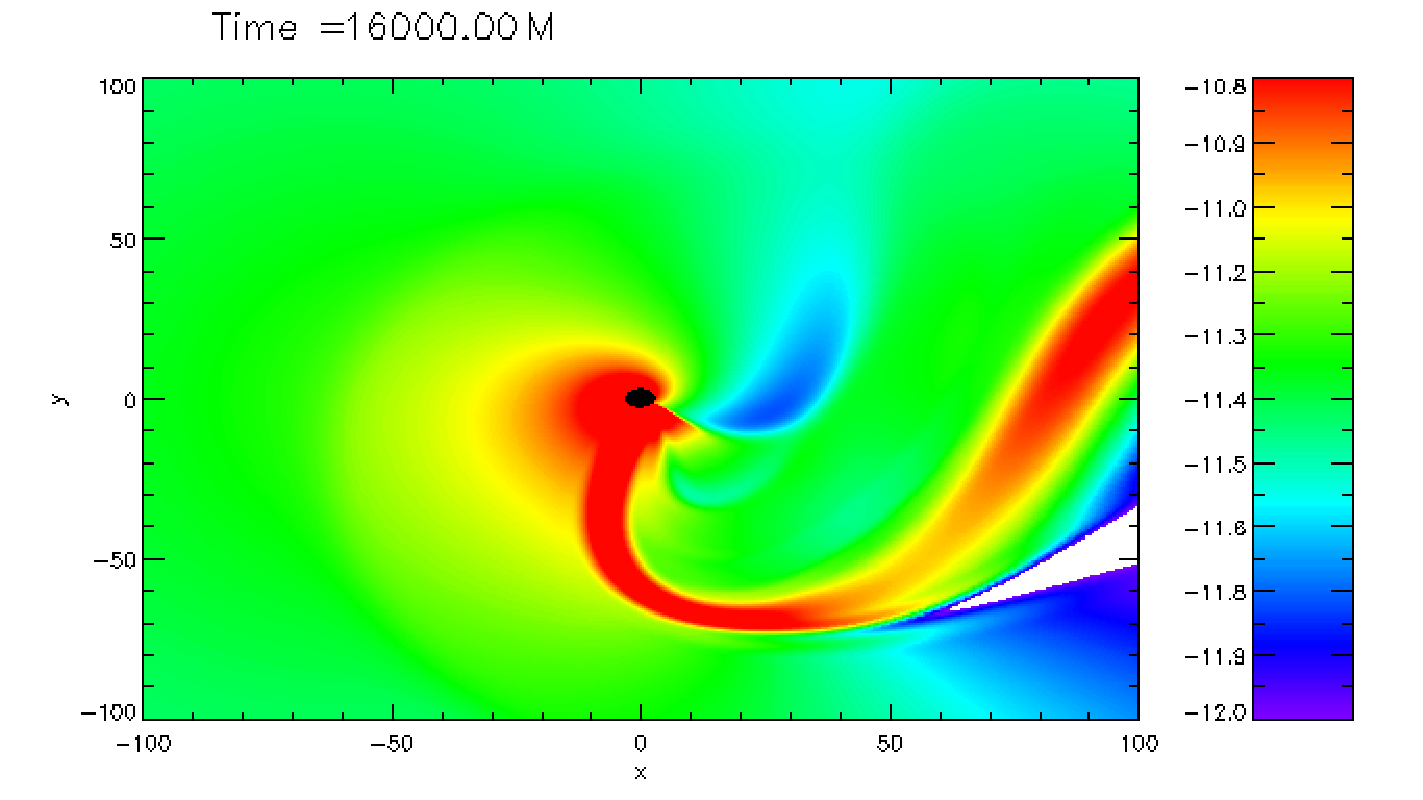}}
{\includegraphics[angle=0,width=8.5cm,height=8.0cm]{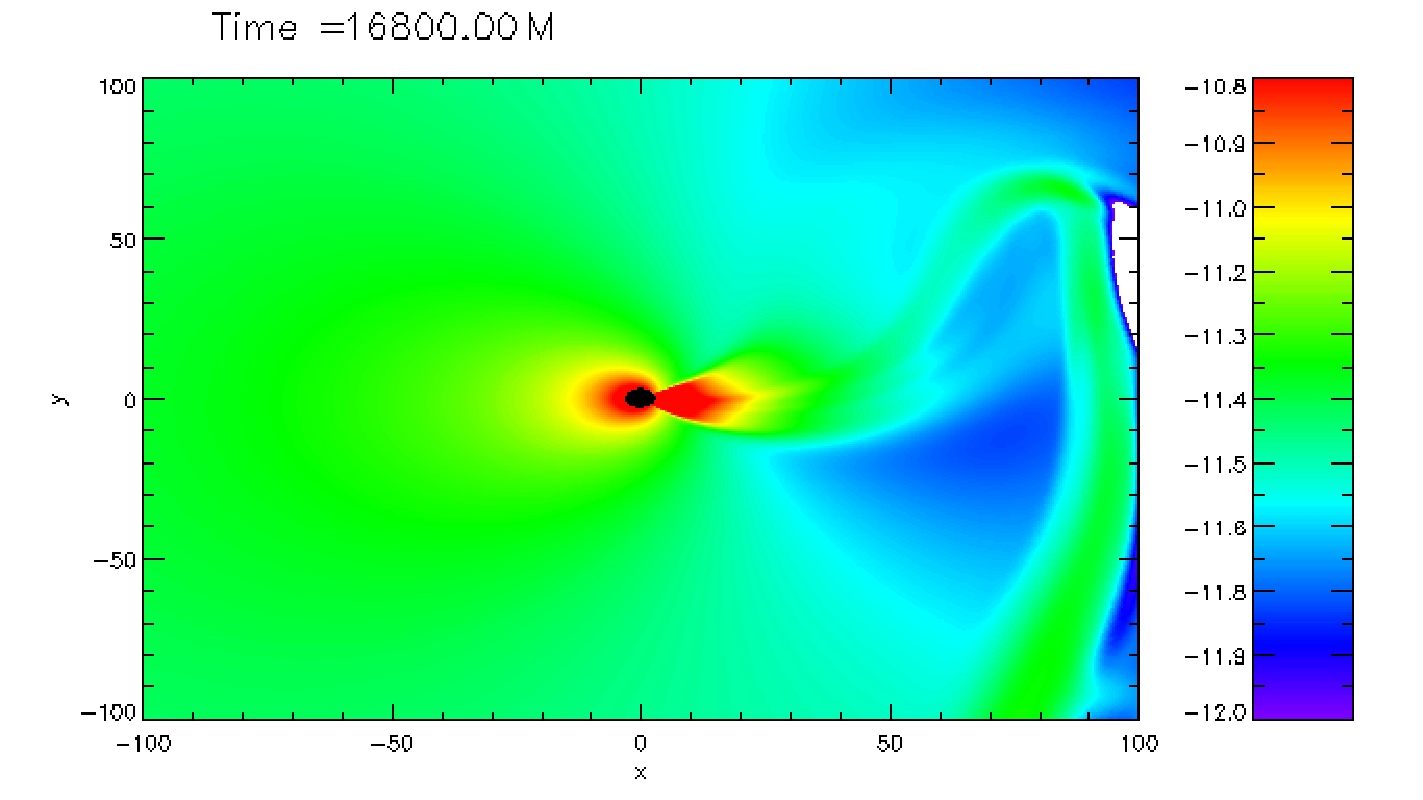}}
{\includegraphics[angle=0,width=8.5cm,height=8.0cm]{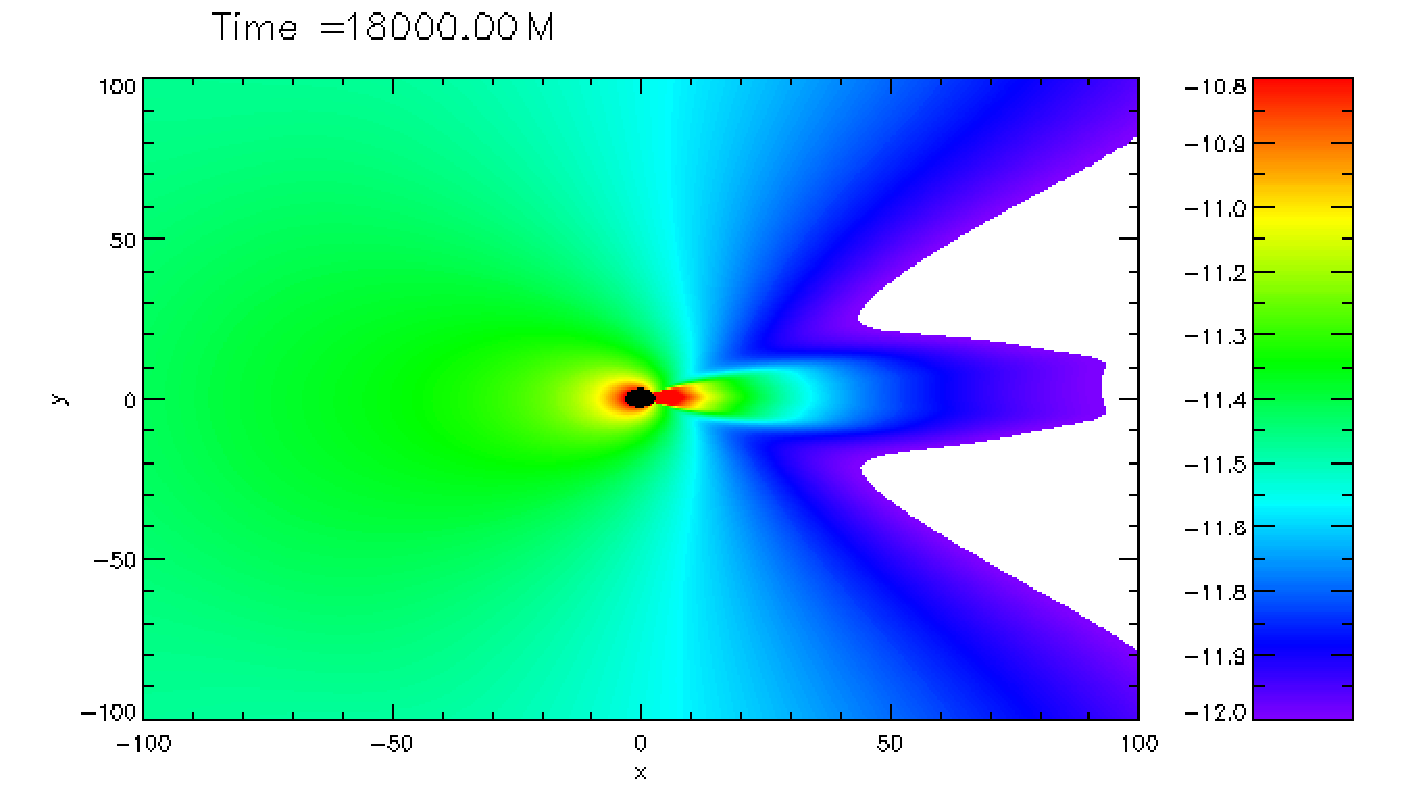}}
\caption{Rest-mass density in cgs units on a logarithmic scale for the
  perturbed Bondi-Hoyle model $\mathtt{p.V10.CS07}$ at four different
  times in a radiation-hydrodynamics evolution. The two rows refer to
  different times of the evolution and white regions correspond to
  densities slightly below the threshold for the colour code at around
  $10^{-12}\,{\rm g/cm}^{-3}$. Note that a highly dynamical transient
  precedes the development of a stationary flow.}
\label{evolution_pmodel}
\end{figure*}

\subsubsection{Perturbed Bondi-Hoyle flow}
\label{Perturbed Bondi-Hoyle flow}
%


As mentioned in Sect.~\ref{sec:Initial_conditions}, in addition to the
standard and stationary Bondi-Hoyle flows, we have also considered
initial conditions that would lead to perturbed Bondi-Hoyle accretion
patterns (we recall that we have tagged these as the ``p-models'').
The rationale behind this choice is that of investigating how the
accretion flows varies when the initial conditions are no longer those
ensuring a stationary flow. At the same time, this allows us to
consider models that have lower temperatures and, consequently, lower
thermal conductivities.

In practice, we trigger the perturbation of the Bondi-Hoyle flow by
acting on the thermodynamic conditions of the fluid in the upstream
region and by producing models with values of the initial temperature
that are typically one order of magnitude smaller than those in
standard Bondi-Hoyle models. Because of the perturbation introduced,
the dynamics of the perturbed models is typically characterized by a
very dynamical phase before quasi-stationarity is reached. However, in
spite of these violent transients, the perturbation introduced does
not destroy the general Bondi-Hoyle pattern, which is recovered
eventually.

Among the perturbed models, $\mathtt{p.V09.CS07}$ has the minimum Mach
number, and is also the only one producing a shock cone that
progressively reverses into the upstream region as a bow shock. The
remaining three models, which all have higher Mach numbers, develop
the usual shock cone downstream of the black hole. This behaviour is
reported in the four panels of Fig.~\ref{evolution_pmodel}, showing
the evolution at different times of the rest-mass density for the
model $\mathtt{p.V10.CS07}$ in a radiation-hydrodynamical
evolution. Note that the accretion cone, that is fully formed at time
$t\sim 3000 \, M$, is highly unstable and it goes through a rapid
sequence of oscillations generating an undulated stream in the
wake. Finally, the system reaches a quasi-equilibrium state
characterized by a reduced shock cone similar to that already
encountered in the dynamics of standard models.

\begin{figure*}
{\includegraphics[angle=0,width=7.5cm,height=7.5cm]{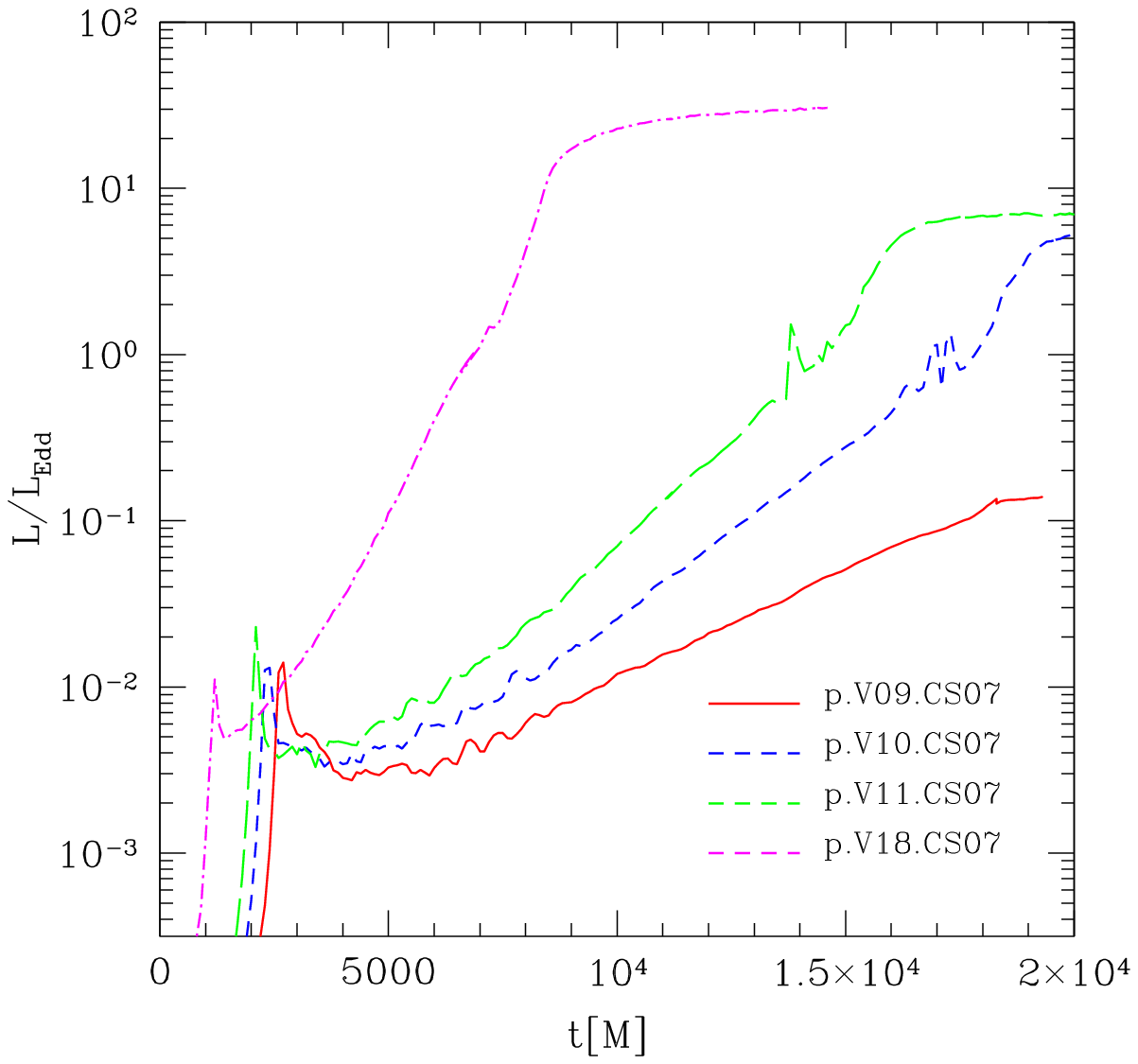}}
\hskip 1.0cm
{\includegraphics[angle=0,width=7.5cm,height=7.5cm]{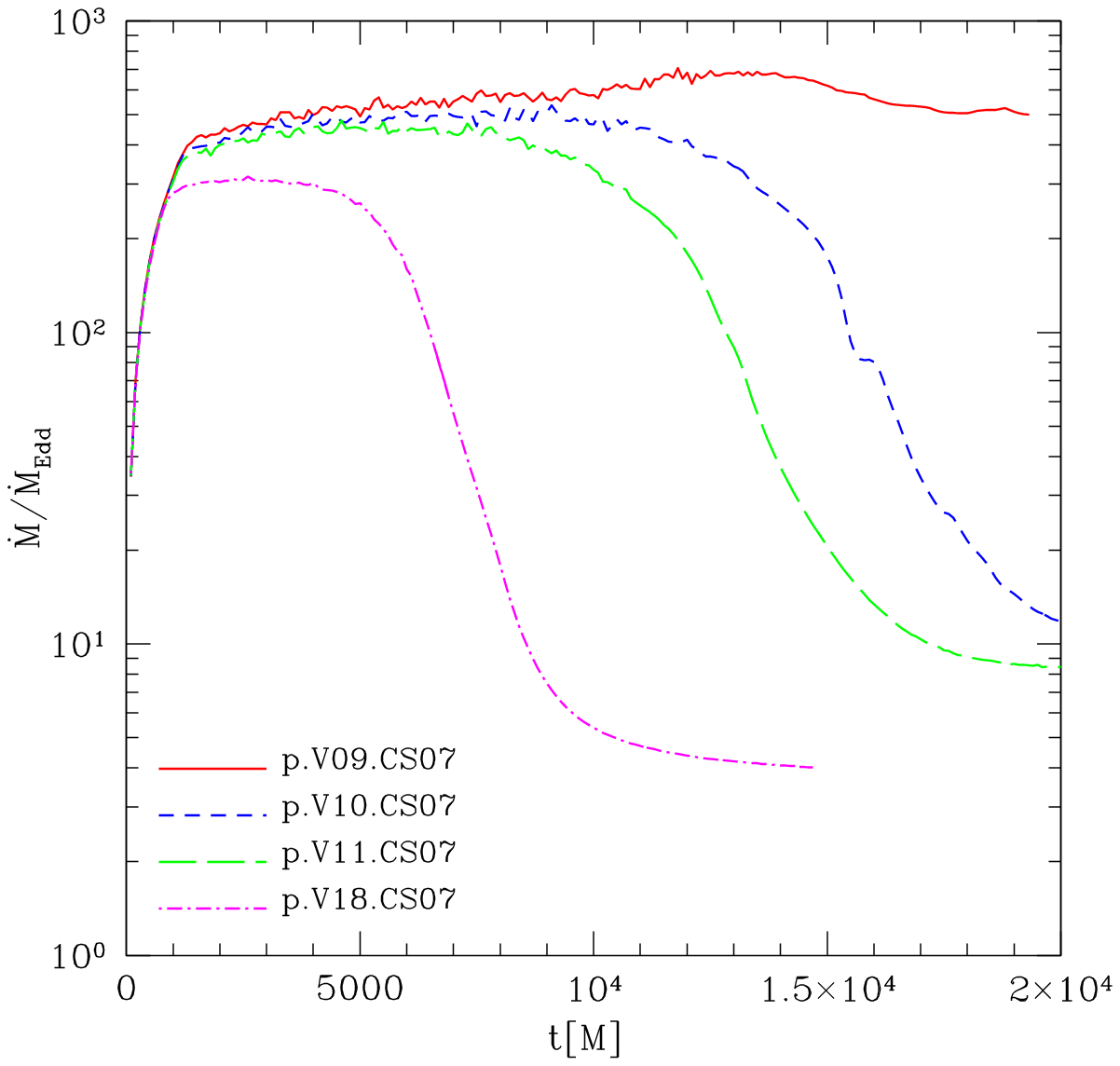}}
\caption{\textit{Left Panel:} Luminosity in Eddington units for the
  perturbed Bondi-Hoyle accretion flows. \textit{Right Panel:} Mass
  accretion rates in Eddington units for the same models in the left
  panel. }
\label{fig:accretion_rates_pmodels}
\end{figure*}

The extraction of the light-curve and the computation of all remaining
quantities follows the same procedure used in the standard models and
we have reported the mass-accretion rates and the light-curves in the
two panels of Fig.~\ref{fig:accretion_rates_pmodels}. Note that the
general features in the light-curves for the standard models are also
present for the perturbed models. In particular, there is an initial
rise in luminosity which corresponds to the formation of the shock
cone. After that, between $t\sim1000 \, M$ and $t\sim2000 \, M$
depending on the model, a peak is produced in the light-curve which is
due to the shock cone changing its geometry to an open
cone. Interestingly, even the efficiency $\eta_{_{\cal BH}}$  of the
perturbed models are very similar to those of the corresponding
standard ones. For the model $\mathtt{p.V11.CS07}$, for instance,
$\eta_{_{\cal BH}}=0.40$, to be compared with $\eta_{_{\cal BH}}=0.38$
of $\mathtt{V11.CS07}$. 

We remark that the fluid temperature within $25 \,M$ from the black
hole decreases more rapidly for high Mach numbers, so that the build
up of the radiation pressure is faster for the highest Mach number. It
should also be noted that while all perturbed models are
radiation-pressure dominated in the upstream region after $t\sim 10000
\, M$, this regime is reached at different times by different
models. Furthermore, even when radiation pressure dominates the
dynamics, there could be isolated portions of the flow where the gas
pressure is not completely negligible. This is the case, for instance,
in the undulated downstream part of the flow, where the ratio of gas
pressure to radiation pressures ratio can be as high as $p/{\cal
  P}_{\rm r}\sim 0.1$.

The dominant role played by the radiation pressure is imprinted on the
accretion rate for the p-model $\mathtt{p.V18.CS07}$, as it is clear
from Fig.~\ref{fig:accretion_rates_pmodels}. This model
features the lowest quasi-equilibrium accretion rate and the highest
luminosity. In general, we have found that the higher the Mach number,
the higher the radiation pressure, and the smaller the average density
around the black hole. The perturbed model $\mathtt{p.V18.CS07}$ shown
in Fig.~\ref{evolution_pmodel}, for instance, has a rest-mass density
which is a factor $100$ smaller than that in model
$\mathtt{p.V09.CS07}$, which has the minimum Mach number among the
perturbed models and the longest relaxation time (\cf
Fig.~\ref{fig:accretion_rates_pmodels}). At the same time, the
accretion rate of $\mathtt{p.V09.CS07}$ does not show the typical
decline up until $t=20000 \, M$, although it is radiation-pressure
dominated everywhere in the numerical domain. It is possible that the
behaviour of model $\mathtt{p.V09.CS07}$ would change, with the
mass-accretion rate decreasing and the luminosity increasing, if the
evolution was carried on a much longer timescale. 

\subsubsection{Spinning black holes}
\label{sec:sbhs}

Although the results presented so far refer to Schwarzschild black
holes, a number of different simulations have been performed also for
spinning black holes, with dimensionless spin parameters ranging
between $0$ and $0.999$. The interest, in these cases, was that of
determining the influence that the black-hole spin may have on the
flow pattern and on the emission properties, for both the classical
Bondi-Hoyle configurations and the perturbed ones.

Overall, the modifications introduced by the black-hole spin are not
particularly large to deserve a dedicated discussion. More
specifically, as far as the dynamics is concerned, we have confirmed
that as the spin of the black hole is increased, the shock cone that
may form in the downstream part of the flow is progressively wrapped
(this was originally pointed out by~\citep{Font1999b}). This
distortion, however, is evident only in the immediate vicinity of the
horizon, and typically below $r\leq 20\,M$. Furthermore, no
significant change has been found, either qualitatively or
quantitatively, in the luminosity, to the point that the light-curves
for different black hole spins overlap to within $1\%$. These results
suggest that if spin-related signatures in the electromagnetic
emission should exist and can be extracted, these will become evident
only when a more sophisticated modelling of the emission processes
(\eg through inverse Compton in a rarefied corona) will be
considered. This will be part of our future work.

\subsubsection{Impact on electromagnetic counterparts 
of supermassive black-hole binaries}
\label{sec:iEMCs}

Considerable attention has been recently dedicated to the possibility
of detecting the electromagnetic counterpart of inspiral and merger of
supermassive binary black holes (SMBBHs) systems.  Such a detection
would not only confirm the gravitational-wave detection and help
localize the source on the sky, but it would also provide a new tool
for addressing a number of astrophysical questions (see,
\eg~\citet{Haiman2009b}). These include the possibility of testing
models of galaxy mergers and clues on the mass distribution of
supermassive black holes (see, \eg~\citet{Sesana2011} and references
therein).

Computing the EM counterpart to the inspiral of such a
binary~\textit{is not} an aspect of our investigation and the physical
conditions considered here badly match those expected in realistic
scenarios describing this process, to which we plan to dedicate a
separate investigation. The results obtained here, however, can serve
to shed some light about a common approximation made in numerical
simulations aimed at estimating the luminosity from binary black hole
mergers ~\citep{Bode:2009mt, Farris:2009mt, Oneill2009, Megevand2009,
  Zanotti2010, Bode2011, Farris2011}. All these works computed the
bremsstrahlung luminosity without taking the back-reaction of the
radiation into account, but rather performing a volume integral of the
bremsstrahlung emissivity.  First efforts to improve this treatment,
but still without a proper radiation transfer, were initiated in
Newtonian physics by~\citet{Corrales2009} by enforcing an isothermal
evolution.

To prove our conjecture that the estimates made so far in terms of the
bremsstrahlung luminosity are optimistic, providing cooling times that
are too short, we have computed the bremsstrahlung-luminosity emitted
in the classical Bondi-Hoyle accretion of model $\mathtt{V09.CS07}$
following the general relativistic prescription adopted by the works
cited above, namely
\be
\label{brem_geo}
L_{{\rm BR}}\simeq 3 \times 10^{78} \int
\left( T^{1/2}\rho^2 \Gamma \sqrt{\gamma}d V \right)  
\left(\frac{M_\odot}{M}\right)
\ \ {\rm erg}/{\rm s} \ ,
\ee
and compared the results obtained using the estimate~(\ref{brem_geo})
with those obtained through our radiative-transfer treatment.  In
addition, we have also considered an alternative calculation in which
an isothermal evolution is enforced, and where it is assumed that all
the changes in the temperature that are due to a local compression are
dissipated as radiation. This idea, proposed in Newtonian framework
by~\citet{Corrales2009}, has been extended to a general-relativistic
context by~\cite{Zanotti2010}, and used also here for comparison.

\begin{figure}
{\includegraphics[angle=0,width=7.5cm,height=7.5cm]{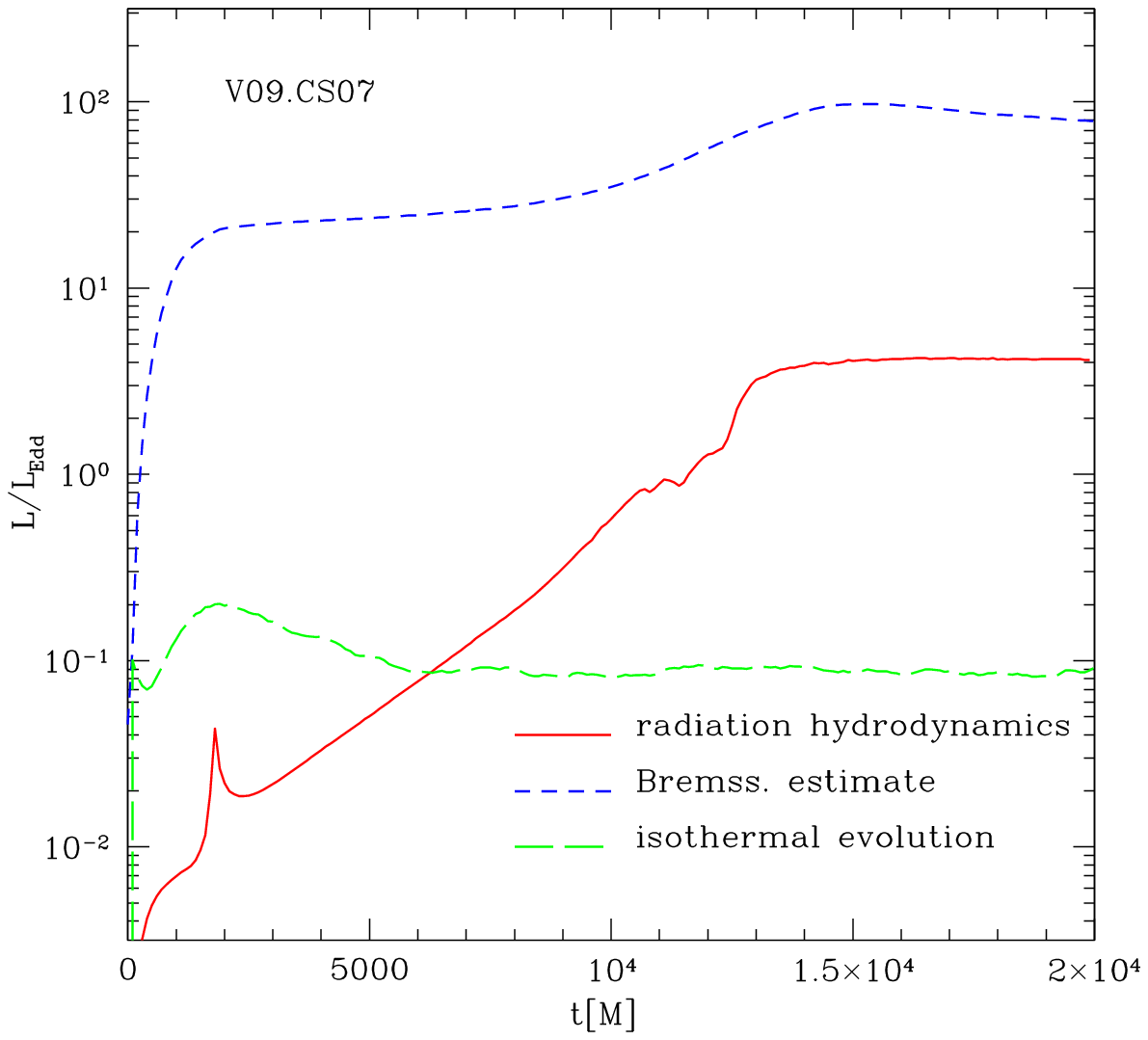}}
\caption{Comparison among light-curves computed with different
  approaches. \textit{Solid red}: luminosity obtained with the full
  radiation-hydrodynamics evolution according to
  Eq.~\eqref{luminosity}. \textit{Dashed blue}: luminosity obtained
  from Eq.~\eqref{brem_geo}.  \textit{Long-dashed green}: luminosity
  obtained through the isothermal evolution approximation. See text
  for more explanations.  }
\label{fig:compare_lum}
\end{figure}

The result of this comparison is shown in Fig.\ref{fig:compare_lum},
where we have reported the three light-curves computed according to
the approaches just described.  When stationarity is reached, \ie at
$t=20000\,M$ we find that $L_{{\rm BR}}/L_{\rm Edd}=78$. This number
should be contrasted with the result obtained through our
self-consistent radiation-hydrodynamics simulations, which instead
indicate $L/L_{\rm Edd}= 4.11$ (\cf
Table~\ref{table:eta}). Interestingly, the luminosity obtained through
the isothermal approximation provides a much smaller value, \ie
$L/L_{\rm Edd}=0.09$.


While this analysis is not exhaustive and has been performed in the
specific scenario of an optically thick 
Bondi-Hoyle accretion, it does point out that the
predictions made using the simplistic estimate of the bremsstrahlung
luminosity via Eq.~(\ref{brem_geo}) provide light-curves that are a
factor $\sim 20$ larger than those obtained with a more rigorous
approach.

\section{Conclusions}
\label{sec:conclusions}

We have implemented and solved in an extension of the \texttt{ECHO}
code~\citep{DelZanna2007} the equations of relativistic radiation
hydrodynamics in the optically thick regime and on a fixed black-hole
spacetime when these equations are written in a conservation
form~\citep{Farris08}. Within a $3+1$ split of spacetime, we have
discretized in time the set of equations with the method of lines and
performed the evolution in time with a second-order modified Euler
scheme. A fifth-order finite-difference algorithm based on an upwind
monotonicity-preserving filter was employed for spatial reconstruction
of primitive variables, whereas a two-wave HLL Riemann solver was used
to ensure the shock-capturing properties. The new scheme has been
successfully validated through a series of tests involving radiative
shock tubes.

As a first application of the new code we have considered the emission
properties of a hot Bondi-Hoyle accretion flow onto a black hole with
the opacity given by Thomson scattering and thermal bremsstrahlung
only. By considering different models with initial temperatures around
$T\sim 10^{10}K$, an ideal-gas EOS with adiabatic index $\gamma=5/3$,
and various sub-sonic and super-sonic regimes, we have found that the
inclusion of radiation drastically alters the well known dynamics of
Bondi-Hoyle flows in all models considered. In particular, the system
quickly enters a radiation-pressure dominated regime, characterized by
mass accretion rates that, once stationarity is reached, decrease by
one or two orders of magnitude with respect to the purely
hydrodynamical evolution. Nevertheless, the measured accretion rates
are found to be always super-Eddington and as high as $\dot M/\dot M_{\rm
  {Edd}}\sim 25$. This is in agreement with the expectation that the
Eddington limit should hold strictly only in spherically symmetric
flows.  In addition, because the effective adiabatic index in the
radiation dominated pressure regime is smaller than the nominal one of
the gas, the radiation can prevent the reversal of the shock cone that
is typical of Bondi-Hoyle flows with low Mach numbers.

By computing the emitted luminosity through a surface integral over
the radiative fluxes at the last optically thick surface, our approach
has allowed the first self-consistent computation of the light-curves
for the Bondi-Hoyle flow, finding luminosities $L/L_{\rm {Edd}} \simeq
1 - 7$. These results have been found to be independent of the initial
conditions chosen for the intensity of the radiation energy density.

In addition to the classical Bondi-Hoyle accretion flows, we also
performed simulations with perturbed setups, by injecting
lower-temperature matter in the upstream region of the flow, which
lead to highly-dynamical transients reminiscent of the flip-flop
instability~\citep{Foglizzo2005}. Although the qualitative evolution
of the accretion flow remains unchanged, the decreased initial
temperature increases the timescale over which the flow becomes
radiation-pressure dominated and the accretion settles in a
quasi-stationary state. In spite of these differences, we have found
that the main features of the Bondi-Hoyle solution, such as the
presence of the shock cone, persist under a wider class of physical
conditions, even in situations departing from stationarity.  Overall,
our results confirm and extend related Newtonian studies, such as
those by~\citet{Kley1995}.

Since we have shown that the luminosity is critically affected by the
evolution of the \emph{coupled} system of hydrodynamic and radiation
equations, significant changes in the luminosities should be expected
in those scenarios which have so far been modeled through a-posteriori
calculations to a purely hydrodynamical evolution.  A first example in
this respect is given by multi-colour black-body spectra, while a
second example is represented by the calculation of electromagnetic
counterpart to the inspiral of supermassive binary black-hole
systems~\citep{Bode:2009mt, Farris:2009mt, Oneill2009, Zanotti2010,
  Bode2011, Farris2011}. Postponing a more detailed calculation of
this process to a future work, we have here shown that the 
calculation of the bremsstrahlung luminosities adopted in the above
works leads to optimistic estimates, which should be regarded
as upper limits.

\section{acknowledgements}
We are grateful to Luca Zampieri for many discussions and Alberto
Sesana for important comments. We wish to thank Nico Budewitz for his
help with the code and the AEI clusters. The computations were
performed on the \textit{datura} and \textit{damiana} clusters at the
AEI and on the IBM/SP6 of CINECA (Italy) through the ``INAF-CINECA''
agreement 2008-2010. This work was supported in part by the DFG grant
SFB/Transregio~7.

\appendix
\section[]{Extended geometrized system of units}
\label{appendixA}

We recall that the definition of geometric units of time and lengths
is obtained by setting the speed of light $c$ and the gravitational
constant $G$ to pure numbers. This implies that seconds and grams of
the $\rm {cgs}$ system can be written as
\begin{eqnarray}
\label{second}
1 {\rm s}&=&2.997924 {\times} 10^{10}\,\left(\frac{1}{c}\right)\, \rm{cm} \,  \\
\label{gram}
1 \rm{g}&=&7.424157{\times} 10^{-29}\,\left(\frac{c^2}{G}\right)\, \rm{cm} \,.
\end{eqnarray}
Within this general setup, a convenient unit of space is still
required. The $\rm{cm}$ is of course a bad choice and the
gravitational radius $r_g=G M/c^2$ is instead chosen. In order for
this new unit to be convenient with respect to the centimeter, the
mass $M$ of the system has to be sufficiently large.  From the
physical value of the solar mass and from (\ref{gram}) we find the
relation between the cgs units and the new unit of length $r_g$
\bea
1\rm{cm}&=&6.772289 {\times} 10^{-6}\,
\left(\frac{M_{\odot}}{M}\right)\, r_g\,, \\
1\rm{s}&=&2.030281 {\times} 10^{5} \,
\left(\frac{1}{c}\right) \left(\frac{M_{\odot}}{M}\right)\,r_g\,, \\
1\rm{g}&=&5.027854{\times} 10^{-34} \,
\left(\frac{c^2}{G}\right) \left(\frac{M_{\odot}}{M}\right)\,  r_g \,.
\eea
It is also useful to write explicitly the conversion of
rest-mass density and luminosity
between the two systems, namely
\bea
\label{conv_rho}
&&\rho_{\rm{cgs}}= 6.1776 {\times} 10^{17}\, 
\left(\frac{G}{c^2}\right)\left(\frac{M_{\odot}}{M}\right)^2\,
\rho_{\rm{geo}} \ , \\
&&L_{\rm cgs}=3.6292\times 10^{59}\,
\left(\frac{G}{c^5}\right)\, L_{\rm{geo}} \ , 
\eea
where $\rho_{\rm{cgs}}$ and $\rho_{\rm{geo}}$ (as well as
$L_{\rm cgs}$ and $L_{\rm{geo}}$)
are the pure numbers expressing
the mass density (as well as the luminosity)
in the ${\rm cgs}$ system and in the geometrized
system, respectively. In the traditional geometrized system $c$ and
$G$ are set equal to unity. However, for specific physical
applications where very low mass densities are encountered, the
corresponding value of $\rho_{\rm{geo}}$ may become prohibitively small.
For this reason, it is convenient to assume a smaller value of $G$,
such as $G=10^{-10}$.

For convenience, we report the Eddington luminosity
and the Thomson scattering opacity of electrons
in geometrized units, namely
\bea
\label{Edd_geo}
L_{\rm{Edd}}&=&3.4636\times 10^{-22} \,
\left(\frac{c^5}{G}\right)
\left(\frac{M}{M_{\odot}}\right)\ , \\
\label{scattering_thompson_e_geo}
\chi^s_e &=& 3.628\times 10^{22}\, G \, \rho_{\rm{geo}}\,  \left(
\frac{M_{\odot}}{M}\right) \ .
\eea

The extension of the geometrized system of units to the temperature
can be obtained by setting to a pure number any physical constant
containing the temperature. In this paper we have chosen to set
$m_p/k_B=1$, where $m_p$ is the mass of the proton, while $k_B$ is the
Boltzmann constant. In this way the temperature is a dimensionless
quantity and the transformation of the temperature from the
dimensionless values to Kelvin is given by
\be
T_K = 1.088\times 10^{13}\,T_{\rm{geo}}  \,.
\ee
In these {\em extended} geometrized units the radiation constant
$a_{\rm rad}=4\sigma/c$ becomes
\be
a_{\rm rad}= 0.191495 \,
\left(\frac{1}{G}\right) \left(\frac{M}{M_{\odot}}\right)^2 \, r_g^{-2} \,.
\ee
%


\bibliographystyle{mn2e}
\bibliography{biblio/aeireferences}

\bsp

\label{lastpage}

\end{document}